\documentclass{aa}  
\usepackage{natbib}
\usepackage{graphicx,xspace}
\usepackage{url}
\usepackage{threeparttable}
\usepackage{amsmath, amssymb}
\usepackage{color}
\usepackage{microtype}
\usepackage{booktabs, makecell}
\usepackage{subfig}
\usepackage{mwe}
\usepackage{lscape}
\usepackage[varg]{txfonts}
\usepackage{longtable}
\usepackage{arydshln}
\usepackage{multicol}
\usepackage{placeins}

\newcommand\fergs{\ensuremath{\mathrm{erg}\,\mathrm{cm}^{-2}\,\mathrm{s}^{-1}}\xspace}

\newcommand\kmps{\ensuremath{\mathrm{km}\,\mathrm{s}^{-1}}\xspace}
\newcommand\cps{\ensuremath{\mathrm{cts}\,\mathrm{s}^{-1}}\xspace}
\newcommand\lx{\ensuremath{\mathrm{erg}\,\mathrm{s}^{-1}}\xspace}

\newcommand{\xmm}{{\it XMM-Newton}\xspace}
\newcommand{\chandra}{{\it Chandra}\xspace}
\newcommand{\srg}{{\it SRG}\xspace}
\newcommand{\mkcflow}{{\tt mkcflow}\xspace}
\newcommand{\eSASS}{{\tt eSASS}\xspace}
\newcommand{\nrta}{{\tt NRTA}\xspace}

\begin{document}

   \title{Old Galactic novae in the eROSITA All Sky Survey}


   \author{Gloria Sala \inst{1,2}
          \and
          Frank Haberl \inst{3}
          \and
          Axel Schwope \inst{4}
          \and
          Dusán Tubín-Arenas\inst{4,5}
           \and
          Elif \c{S}afak\inst{1}
           \and
          Chandreyee Maitra\inst{3}
           \and
          Jochen Greiner\inst{3}
          }  

   \institute{Departament de Física, EEBE, Universitat Politècnica de Catalunya, c/Eduard Maristany 16, 08019, Barcelona, Spain\\
              \email{gloria.sala@upc.edu}
         \and
             Institut d'Estudis Espacials de Catalunya (IEEC), c/ Esteve Terradas 1, 08060 Castelldefels (Barcelona), Spain 
         \and
             Max-Planck-Institut für extraterrestrische Physik (MPE), Gie{\ss}enbachstra{\ss}e 1, 85748 Garching, Germany
        \and
             Leibniz-Institut für Astrophysik Potsdam (AIP), An der Sternwarte 16, 14482 Potsdam, Germany
        \and
         Potsdam University, Institute for Physics and Astronomy, Karl-Liebknecht-Straße 24/25, 14476 Potsdam, Germany
             }
   \date{}

 
  \abstract
   {Nova explosions occur on accreting white dwarfs. A thermonuclear runaway in the H-rich accreted envelope causes its ejection without destroying the white dwarf, and an increase in the luminosity by several magnitudes. Accretion is re-established some time after the explosion. The explosion of the nova itself is expected to affect the mass-transfer rate from the secondary and the accretion rate, but these effects have been little explored observationally. Most novae are observed only in outburst and the properties of the host systems are unknown.}
   {X-ray observations of novae happen mostly during outburst; only a few have been the target of dedicated X-ray observations years or decades after outburst. However, the X-ray emission long after the outburst provides a powerful diagnostics of the accretion rate and the possible magnetic nature of the white dwarf.}
   {We have explored the first two years of the SRG/eROSITA All Sky Survey (eRASS) for X-ray sources correlated with Galactic historical novae. We present the first population study of nova hosting systems in X-rays, focus on the evolution of accretion rate as a function of the time since last outburst, and look for new candidates for magnetic systems.}
   {In total, 32 X-ray counterparts of novae are found in the western Galactic hemisphere. Combined with 53 nova detections published for the eastern hemisphere, the fraction of X-ray detected novae in quiescence is 18\% of the Galactic novae. We have, for the first time, enough statistics to  observationally determine the evolution of accretion rate as a function of time since the last nova outburst for a time span of 120 years. The results confirm that magnetic systems remain systematically at higher fluxes and that accretion is enhanced during the first years after the outburst, as predicted theoretically. We also identify new IP candidates in AT Cnc and RR Cha.}
   {}


   \maketitle
%
\section{Introduction}

Nova outbursts are thermonuclear explosions in the accreted envelope of a white dwarf (WD). Most novae occur in Cataclysmic Variables (CVs), binary stellar systems with a low-mass donor transferring material to the WD via Roche lobe overflow and forming an accretion disk. In some systems, the WD accretes directly from the dense wind of a red giant companion in a wider binary system. The mechanism of mass accretion (disk or wind) does not affect the final outcome of the WD's accreted envelope: the H-rich material freshly accumulated under extreme conditions will ignite, triggering a thermonuclear runaway that ultimately leads to the ejection of (most of) the accreted envelope. Novae eject $10^{-7}-10^{-4}M_{\odot}$ of material enriched with freshly synthesized elements at velocities of several thousands \kmps \citep{1998ApJ...494..680J, 2016PASP..128e1001S}.

Nova explosions are relatively frequent events: the estimated Galactic rate is $50(^{+31}_{-23})$ novae per year \citep{2017ApJ...834..196S}. However, only about one-fifth of these events are observable, and most of them are lost because of the extinction. In contrast to Type Ia supernovae (SNIa), neither the WD nor the binary system is disrupted during a nova outburst. Novae constitute indeed recurrent phenomena, with expected periodicity of $10^4-10^5$~years in most cases. The most interesting subclass are recurrent novae (RNe), defined by more than one \emph{observed} outburst, that is, periodicity of 1 to 100 years. Only 10 objects belong to this class \citep{2021gacv.workE..44D}. The higher recurrence frequency points to a larger WD mass, close to the Chandrasekhar limit, and a high accretion rate, making them potential candidates for SNIa progenitors \citep{2003MSAIS...3..129S, 2012BASI...40..393K}.

While the general picture is well understood, the details of the processes are full of unknowns. As an example, the initial fireball phase of the explosion predicted from theory since decades was only recently detected for the first time thanks to eROSITA \citep{2022Natur.605..248K}. Another point of interest and focus of current debates in the field is the contribution of novae to the Lithium present in the Galaxy \citep[after the first detection of Li from Nova Del 2013 by ][]{2015Natur.518..381T}, or the mechanisms powering the Very High Energy Emission \citep{2010Sci...329..817A}, probing the high energetic shocks and particle acceleration occurring in the ejecta already predicted some years before \citep{2007ApJ...663L.101T}. 

The X-ray band is one of the most important diagnosis for novae. In outburst, nova X-rays arise in the form of supersoft emission from the direct H-burning on the white dwarf surface, and on a wider X-ray band extending to harder energies from the expanding, shocked ejecta. In addition, the accretion-powered X-ray emission, arising sooner or later after the outburst, probe the reestablished accretion onto the white dwarf. High-resolution spectroscopy allows us to
disentangle the energetics and composition of the atmosphere and the ejecta, while the variability of the X-ray emission often points out the asymmetries in the system \citep{2008ApJ...675L..93S}. X-ray emission is also a powerful diagnostic tool for the accretion status of novae after outburst, pointing out the reformation or status of the disk after the explosion \citep{2002Sci...298..393H, 2012ApJ...745...43N}. 

CVs are the host systems of most nova explosions. The formation of the WD, shrinking to a smaller size, induces fast rotation to the final object (by conservation of angular momentum) and, in some cases, the resulting dynamo powers a strong magnetic field. The companion (donor) star fills its own Roche-Lobe and transfers matter from its outer layers to the WD through the inner Lagrangian point, forming an accretion disk around the WD. The matter in the accretion disk gets hot and bright due to the viscosity in the gas, thus radiating energy and losing angular momentum, migrating to inner radii in the disk until it finally accretes onto the white dwarf surface. Depending on the strength of the magnetic field, the accretion disk might be truncated before the material reaches the white dwarf surface (intermediate polar system, IP) or, if the magnetic field is strong enough, it may completely disrupt the accretion disk (polar). In all magnetic systems, the accretion stream is funnelled to the white dwarf magnetic poles. The supersonic accreting material becomes subsonic close to the white dwarf surface, creating a shock that ionizes the plasma to X-ray emitting temperatures \citep[see][for a review]{2017PASP..129f2001M}. 

While CVs are relatively faint and thus its population is best known only for the solar neighbourhood \citep[see volume-complete population study up to 150 pc by ][]{2020MNRAS.494.3799P}, the nova outburst increases the brightness of the system by 7-8 magnitudes. That makes nova events detectable at any distance in the Galaxy and even at extragalactic distances \citep{2019enhp.book.....S}. In most cases, however, after the nova outburst is over, the underlying CV remains forgotten or even unreachable. The number of nova progenitors known and characterized is limited \citep{2012ApJ...746...61D, 2014ApJS..213...10W, 2016ApJ...817..143W}. Accretion is however observed to be reestablished in several cases soon after the outburst, and this accreting CV can power an X-ray source with luminosities expected in the range $10^{33}-10^{35}$\,\lx. The accreting column emits as a shocked thermal plasma at typical temperatures of some keV for magnetic novae, while in case of non-magnetic, the equatorial accretion boundary layer powers a fainter and relatively colder X-ray spectrum. In both cases, however, the spectral emission ranges up to several keV, with interstellar absorption only affecting the softer part of the spectrum but not completely hiding the source. Therefore, X-ray spectra of the host CV provides crucial information for characterizing the system, including the magnetic nature of the WD, the accretion rate, the mass of the WD, and the abundances of the accreted material \citep{2017PASP..129f2001M}. The possibility of nova outbursts occurring on magnetic WDs was questioned in the past \citep{1988ApJ...330..264L}. It was thanks to the detection of coherent modulation in the X-ray data that GK Per (Nova Per 1901) was identified as an intermediate polar \citep{1985MNRAS.212..917W}.  At the same time, Nova Cyg 1975 was recognized as occurring in a classical polar, V1500 Cyg \citep{stockman+88}. Since then, a few more novae have been recorded on IP candidates:  V1425~Aql \citep{1998MNRAS.293..145R}, V2487~Oph \citep{2002Sci...298..393H}, V4743~Sgr \citep{2006AJ....132..608K, 2016MNRAS.460.2744Z, 2018MNRAS.480.4489Z}, V4745~Sgr  \citep{2006MNRAS.371..459D}, V597~Pup \citep{2009MNRAS.397..979W}, Nova~M31N~2007-12b \citep{2011A&A...531A..22P}, V2491~Cyg \citep{2015ApJ...807...61Z}, HZ~Pup \citep{1997ASPC..121..679A,2020A&A...639A..17W}, V1674~Her \citep{2021ApJ...922L..42D}, and V407~Lup \citep{2024MNRAS.533.1541O}. 

During the ROSAT All Sky Survey (RASS), the positions of 81 old novae were examined, and only 11 of them (14\%) were identified as accreting systems \citep{2001A&A...373..542O}. This study was the last systematic investigation of old novae in the X-ray band. Since the completion of the RASS, more than 200 new nova events have been recorded. More recently, between 2019 and 2022, the eROSITA All Sky Survey has provided a great opportunity to assess the accretion status of all systems that hosted a nova explosion. 

Here, we present the population of old-nova X-ray counterparts detected in the first two years of the eROSITA All Sky Survey \citep{2021A&A...647A...1P}. We have correlated a cataloge of all nova events recorded in the Galaxy with the German half of eRASS:4, as well as with the individual eRASS1, eRASS2, eRASS3, and eRASS4. Results from eRASS5 are also included in the present work if the source location was covered by this last partial survey. Considering all historical novae, the evolution of the accretion rate with the age of the nova (years since outburst) can be assessed, thus testing the effect of the nova outburst in the immediate years after the explosion and the long-debated hibernation scenario. In addition, we aim at the identification of new candidates to magnetic systems that have hosted historical nova outbursts. The interest is bidirectional: (a) what is the effect of the magnetic nature of the host on the nova explosion? With just a few examples of known magnetic CVs hosting novae, and given that each nova explosion has its differential characteristics, the population is too small to identify common patterns that the magnetic nature of the CV could imprint on the nova explosion. However, we often overlook the magnetic nature of the system for most nova hosts. Therefore, it is of great interest to identify new magnetic nova progenitors and characterize them, including determining the WD mass, magnetic field, and accretion rate. And (b), what is the effect of the nova explosion on the accretion evolution of the host system? In the classification of CVs as non-magnetic (Dwarf Novae) and magnetic (IPs and Polars), there are some exceptional cases of systems exhibiting dual behavior, i.e., systems with IP properties that also display dwarf nova outbursts. \citet{2017A&A...602A.102H} reviewed the disk instability model (DIM) to account for the observed dwarf nova outbursts in IPs, concluding that this anomalous behavior could be related to a historical nova explosion in the system. 

In this paper, we present the methods used for the data extraction and analysis in Section~\ref{section_methods}; the global population properties and evolution of X-ray flux over decades are presented in Section~\ref{section_population}, while Sections~\ref{section_new_sources} and \ref{section_old_sources} report the spectral analyses of individual objects with enough statistics. Sections~\ref{section_discussion} and ~\ref{section_summary} provide final discussion and summary.

\section{Methods}
\label{section_methods}

A list of 536 Galactic novae from the XVII century up to 2022 was obtained from the list maintained by Bill Gray\footnote{https://projectpluto.com/galnovae/galnovae.htm}. First inspection of the list showed that several Mira stars, Low- and High-Mass X-ray Binaries, transients and some other variables were included in the list, probably because they were wrongly identified as novae at some point in the literature. We thus cleaned the list from false-novae. Since our main aim is the study of the accretion state in the nova-hosting cataclysmic variables, we also removed symbiotic systems from the sample, since the accretion physics and their X-ray emission is quite different from CVs. We note however that the Symbiotic nova RR~Tel is detected by eROSITA (see Saeedi~et~al., in prep.). The resulting final list contains 467 novae, with outbursts registered between the XVII century and the year 2022. This final list was correlated with a search radius of 10\arcsec\ with the catalogues for eRASS1, eRASS2, eRASS3, eRASS4 and eRASS5 (individual eRASS surveys) as well as with the merged eRASS:3 and eRASS:4 catalogues (the stack of the first three and four complete all-sky surveys). All products were produced with the same eSASS version, calibration, boresight, etc. (configuration c030). For the non-detected novae, we computed upper limits using the methods described in \citet{2024A&A...682A..35T}. 

We have checked that the normalized separation between the optical nova position and its X-ray counterparts follow a Rayleigh distribution, as expected, thus confirming the identifications. A few detections of four sources fall at sigma larger than three. However, they all correspond to bright sources with small positional error and spectral data that confirm the identification. In particular, RR Pic and CP Pup have large separation sigma only in some of the observations (but not all), and their eROSITA spectra shows little variations and is compatible with the spectral results from previous X-ray observations (see \ref{section_old_sources}); the other two sources with large sigma separation are two novae in outburst (V1710 Sco, YZ Ret), affected by pile-up and not included in further analysis in this work. 

For historical novae that belong to the eastern Galactic hemisphere and were not detected and reported in \cite{2021AstL...47..587G}, we provide upper flux limits based on their mirrored position on the western half of the sky. Since the exposure time map is symmetric in ecliptic coordinates due to the scanning strategy of eROSITA, we mirror the ecliptic coordinates of the eastern non-detected novae on the sky, and we compute upper limits at the position of these mirrored ecliptic coordinates. We note that by construction, these new mirrored coordinates will belong to the western Galactic hemisphere and will be observed by eROSITA with the same exposure time as the original coordinates. We confirm that no nearby sources could potentially contaminate the upper limit calculation at the mirrored positions. Thus, in Section~\ref{section_population} we include upper limits for sources in the eastern Galactic hemisphere based on the western Galactic hemisphere data.

Spectra for sources with more than 50 counts were extracted for the individual eRASS and for the merged eRASS:4. Uncertainties quoted in the text are at 90\% confidence, unless otherwise specified. Spectral fitting was done using XSPEC from the HEASOFT package and using {\it cstat} statistics. Distances to the correlated novae are listed in Table~\ref{tab:summary} with their corresponding reference, in most cases from  Gaia DR3 parallaxes \citep{2023AJ....165..163C}. The interstellar hydrogen column in the direction of each source was obtained for each individual detected source considering its position and distance using the 3DNH-tool\footnote{http://astro.uni-tuebingen.de/nh3d/nhtool} developed by \citet{2024arXiv240303127D}; for the non-detected sources we use the HEASARC N$_H$ tool \citep{2016A&A...594A.116H}.

\section{Population properties}
\label{section_population}

In total, 30 eROSITA sources in the German half of data-rights were found to correlate with novae. Table~\ref{tab:summary} provides a summary with the list of all the detected novae, their corresponding outburst year, the absorption column and the distance. The third column of that Table indicates whether the source was detected in the merged eRASS:4 (sm04), or in the merged eRASS:3 (sm03) with no detection in eRASS:4. If it was only detected in an individual eRASS it is indicated with the label (em0\#), where \# is the eRASS number.

A total of 17 old novae were detected in X-rays for the first time out-of-outburst. Another nine novae had previous X-ray detections out-of-outburst. In addition, four of the novae were detected in outburst or shortly after it (YZ Ret 2020, V1706 Sco 2019, V1708 Sco 2020, and V1710 Sco 2021). This represents an increase by a factor of three with respect to the last similar work with ROSAT \citep{2001A&A...373..542O}.

Here we present however the eRASS:4 results corresponding to the western Galactic hemisphere, and most importantly, with a large fraction of it in the southern hemisphere. For historical reasons, there is a clear bias of nova outbursts detected most frequently in the northern hemisphere, and thus our half-sky is not half of the historical population, but less. Indeed, \citet{2021AstL...47..587G} reported the historical novae detected by eROSITA in eRASS:3 in the eastern Galactic hemisphere, with Russian data rights, and they reported the detection of 52 historical novae in X-rays. To provide a complete population study of the eROSITA detected novae, we include results from \citet{2021AstL...47..587G} in several of the population results reported in this section.

\begin{figure}
    \includegraphics[width=0.49\textwidth]{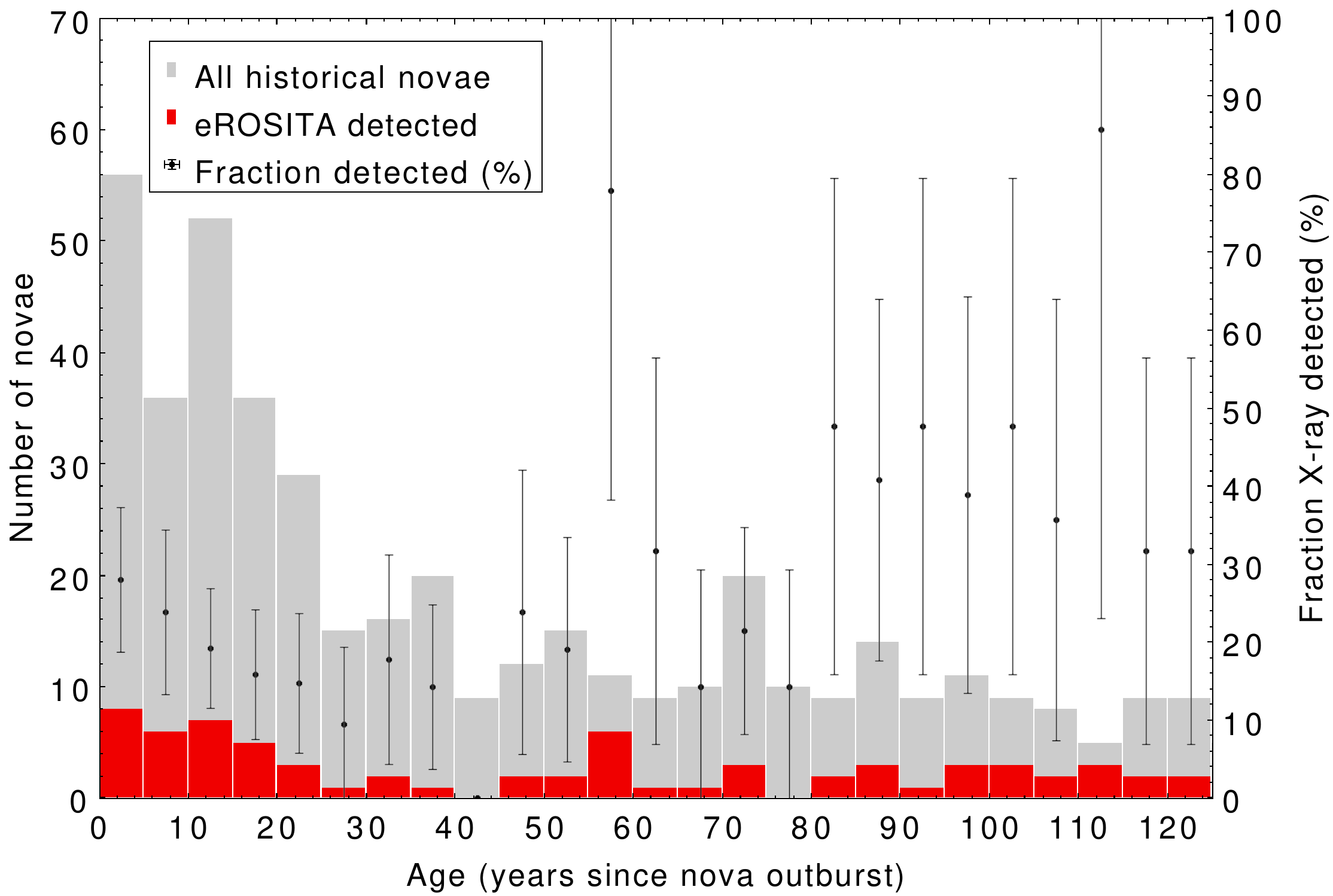}
       \caption{Total number of historical novae (gray bars), of novae detected by eROSITA (red bars) and fraction of eROSITA X-ray detected novae (black bullets, auxiliar right vertical axis) as a function of years since last nova outburst. eROSITA detections include eRASS:4 DE (this work) as well as detections reported by \cite{2021AstL...47..587G}.}
    \label{histogram_all_x_frac}
\end{figure}

Considering the whole sky population, the total number of detected historical novae (not including those detected in outburst) is 83, which represents a fraction of 18\% of our full list of old novae. In Fig.~\ref{histogram_all_x_frac}, the total number of novae, the number of eROSITA detections and the fraction it represents is shown as a function of age, i.e., years since nova outburst. The uncertainty in the fraction of detected novae was calculated as $(XRAY/ALL)\times \sqrt{(1./ALL + 1./XRAY)}$ where XRAY is the number of X-ray detected novae and ALL the total number of historical novae for each bin of the histogram. 

\begin{figure}
    \includegraphics[width=0.5\textwidth]{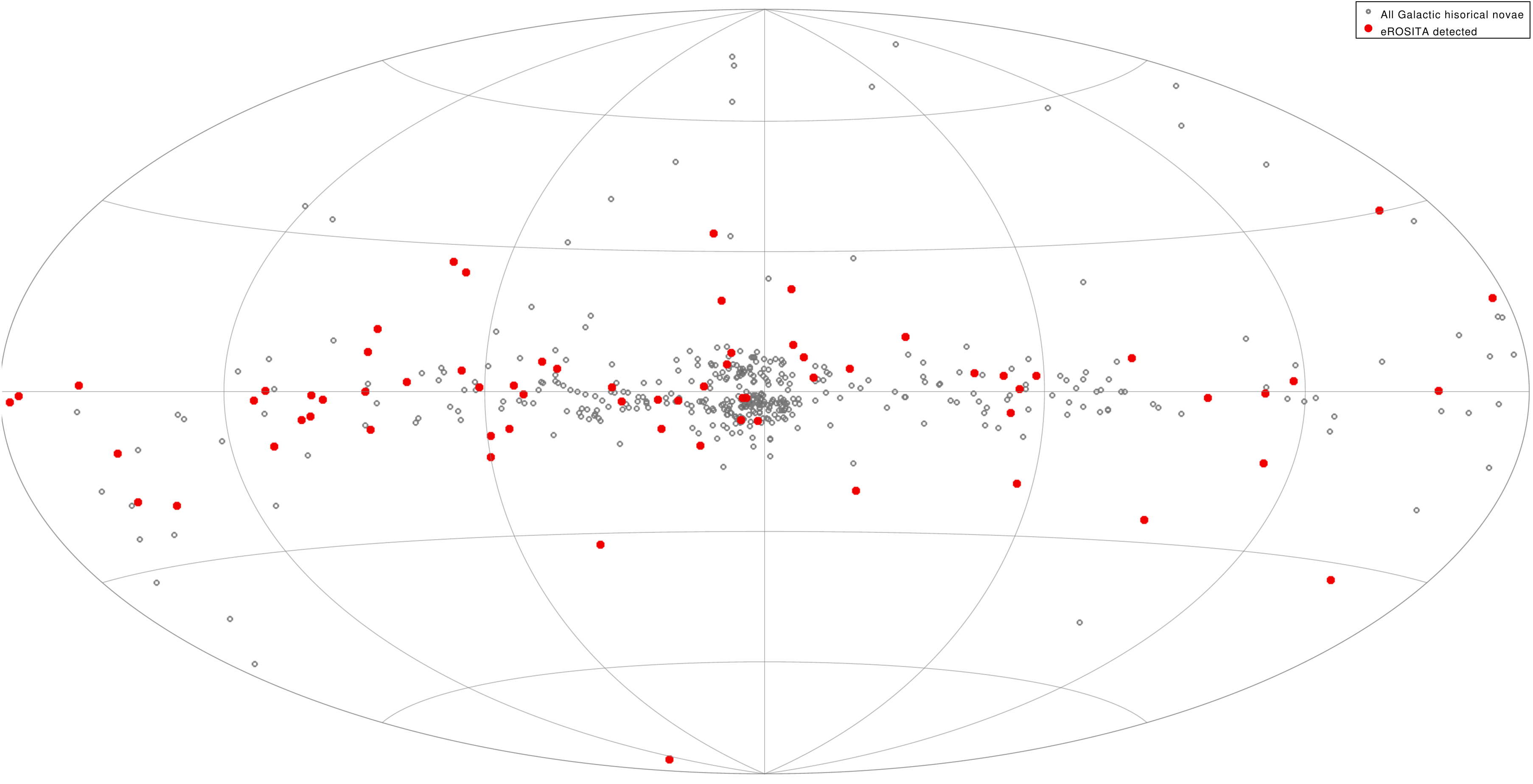}\hfill
       \caption{Galactic distribution of all historical novae (gray) and the novae detected in X-rays by eROSITA (red). eROSITA detections include eRASS:4 DE (this work) as well as detections reported by \cite{2021AstL...47..587G}.}
    \label{aitoff}
\end{figure}

\begin{figure}
    \includegraphics[width=0.5\textwidth]{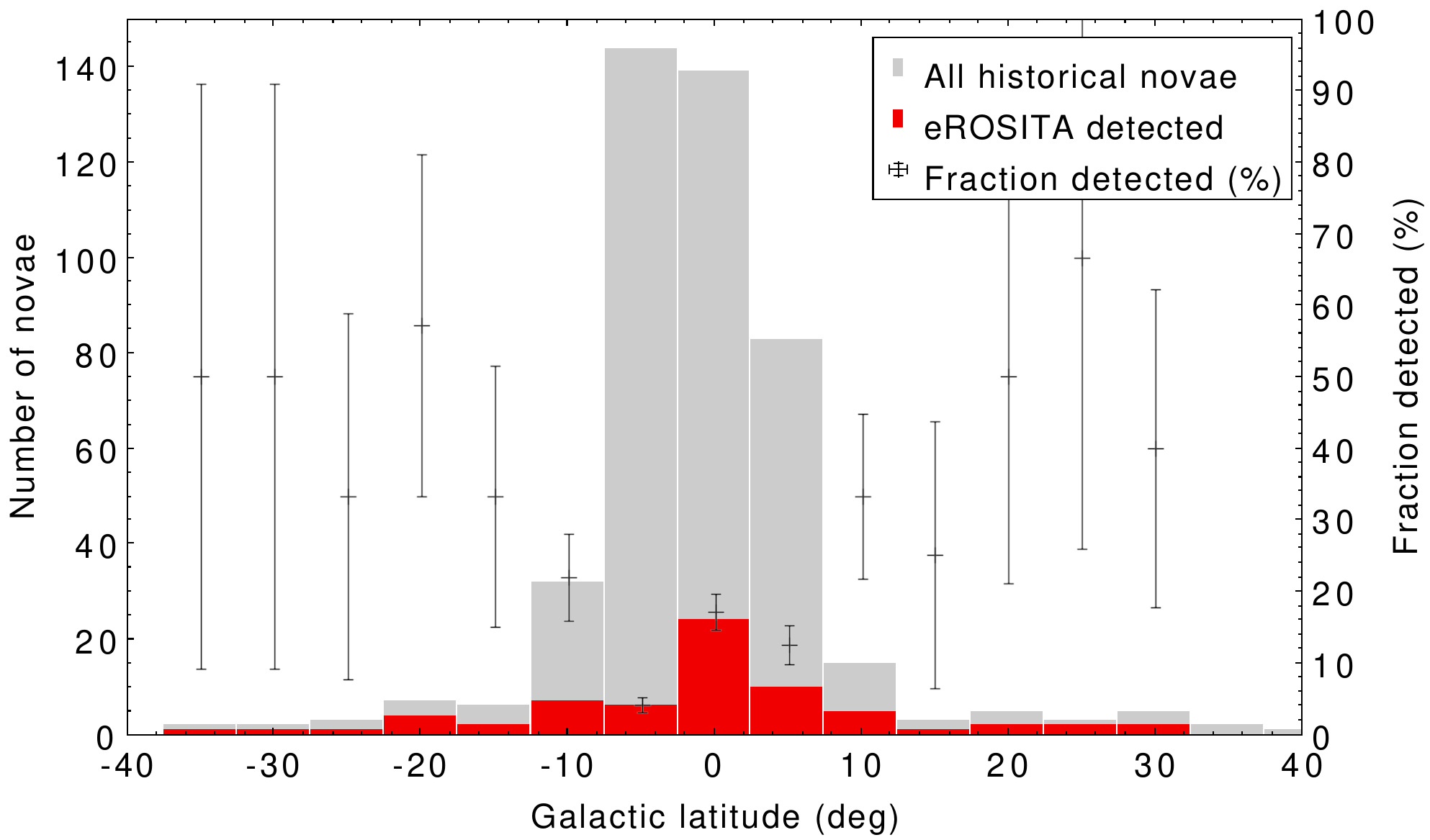}\hfill
    \includegraphics[width=0.5\textwidth]{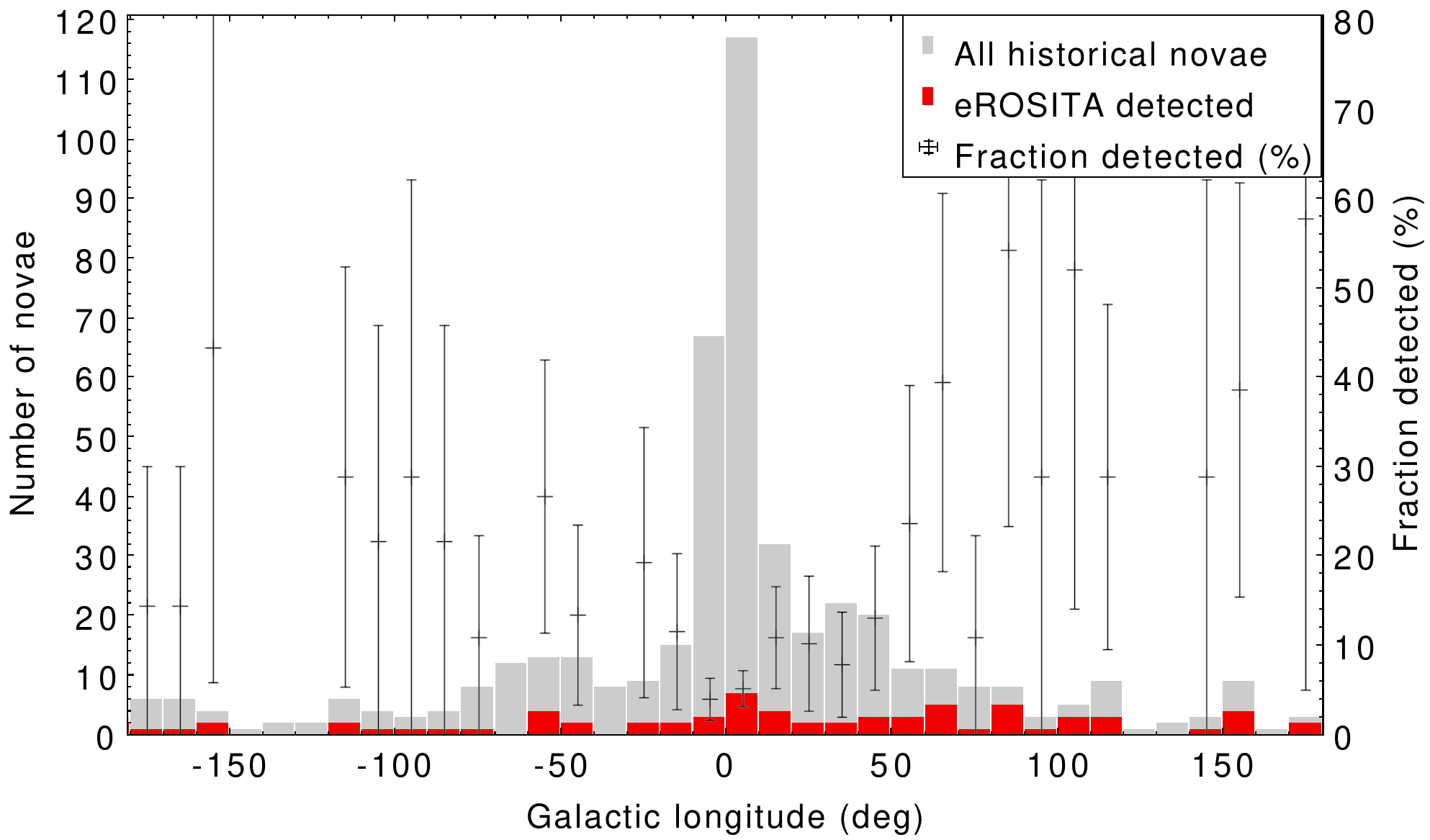}\hfill
       \caption{Total number of historical novae (gray bars), of novae detected by eROSITA (turquoise bars) and fraction of eROSITA X-ray detected novae (in black) as a function of Galactic latitude (upper panel) and longitude (lower panel). eROSITA detections include eROSITA\_DE eRASS:4 (this work) as well as detections reported by \cite{2021AstL...47..587G}.
       }
    \label{gallatlong}
\end{figure}

The location of all the eROSITA detections of historical novae is shown in Fig.~\ref{aitoff}, where also the location of the whole sample of historical novae included is shown. The distribution of all historical novae follows stellar density in the Galaxy, with novae more frequently found in the Galactic plane and towards the Galactic Centre. Inspection of the sky distribution of X-ray detections, however, give a clear impression of deficit of X-ray detected novae towards the Galactic center, most probably due to interstellar absorption. This effect is confirmed in Fig.~\ref{gallatlong} where the distribution of total novae and eROSITA detected ones is shown for Galactic latitude and longitude, as well as the fraction of detections.

\begin{figure*}
    \includegraphics[width=0.99\textwidth]{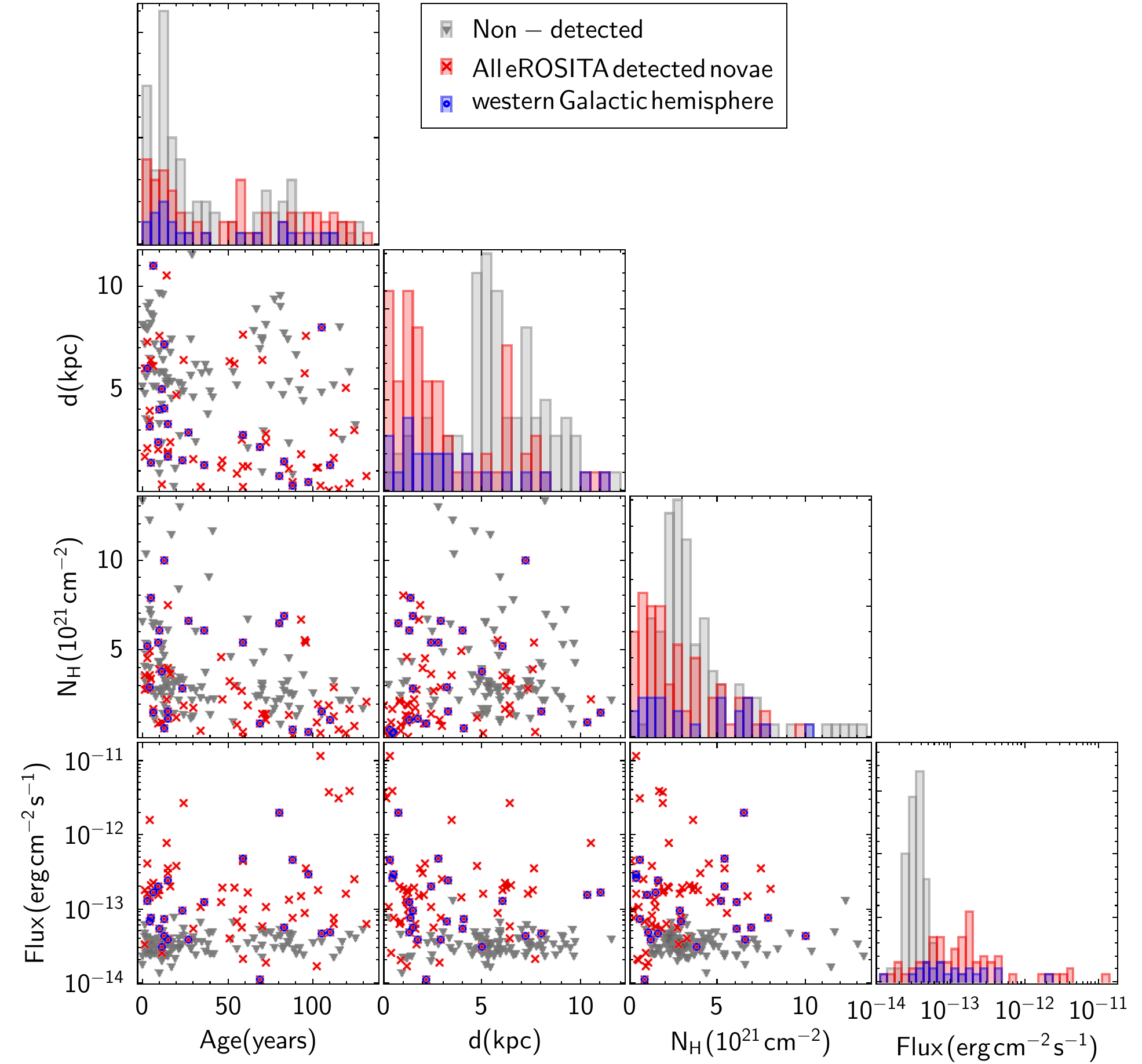}\hfill
       \caption{All sky distributions of all novae detected by eROSITA (red), only those from the western Galactic hemisphere (blue), and upper limits for all non-detected novae in both hemispheres (gray): distribution of age, distance, interstellar hydrogen column and X-ray flux (0.2-2.3 keV) are shown. Whole sky detected novae and upper limits include eastern Galactic hemisphere results reported in \cite{2021AstL...47..587G}.}
    \label{corner}
\end{figure*}

To further understand the distribution of the eROSITA sample of novae in the Galaxy,  we show in Fig.~\ref{corner} the distribution in age (as years since last nova outburst), distance, interstellar hydrogen column, and unabsorbed X-ray fluxes. We show both the novae detected in the western Galactic hemisphere in eRASS:4 and presented for the first time in this work (see also Table~\ref{tab:summary}) as well as the distribution for all the eROSITA detected novae, for the whole sky, including those reported for the eastern Galactic hemisphere by \cite{2021AstL...47..587G}. The unabsorbed X-ray fluxes shown in Fig.~\ref{corner} have been determined assuming a 10 keV thermal plasma emission for the novae of the present work. We have checked that the precise value of the temperature (in the range 5$-$15 keV) does not affect significantly the flux within statistical uncertainties. For the novae reported in \cite{2021AstL...47..587G} and included in the figures of the present work, only unabsorbed luminosities are reported by the authors. The distances assumed for each nova by the authors are also reported in  \cite{2021AstL...47..587G} and come from different sources, some of them known to be out-dated. Therefore, we have corrected for the distances to some objects in \cite{2021AstL...47..587G} to recalculate the luminosity: we have first used the  luminosity and distance reported by  \cite{2021AstL...47..587G} for each source to determine the unabsorbed flux, and then we have recalculated the unabsorbed luminosities with distances from \citet{2021AJ....161..147B}.

Even if the statistics numbers are low, Fig.~\ref{corner} is illustrative to understand the distribution of detected vs non-detected novae. In the distance distribution, it is clear that most detected novae are at distances shorter than 3 kpc, while the novae non detected by eROSITA peak at distances close to the Galactic Center, where historical novae cluster, as can be seen in Fig.~\ref{aitoff}. The absorbing column clearly also plays a role in the detection or non-detection of the objects, with a clear peak of non-detected objects just at values of N$_H \sim (3-5)\times 10^{21}$~cm$^{-2}$ and most detected novae at N$_H$ smaller than $2\times 10^{21}$~cm$^{-2}$. Finally, the upper-limits indicated for the non-detected novae in the flux panel show that most novae with fluxes smaller than $6\times10^{-14}\fergs$  are non detected, although in some cases, longer exposure times in that particular position have led to detection of sources with similar fluxes.

\begin{figure*}
    \includegraphics[width=1.0\textwidth]{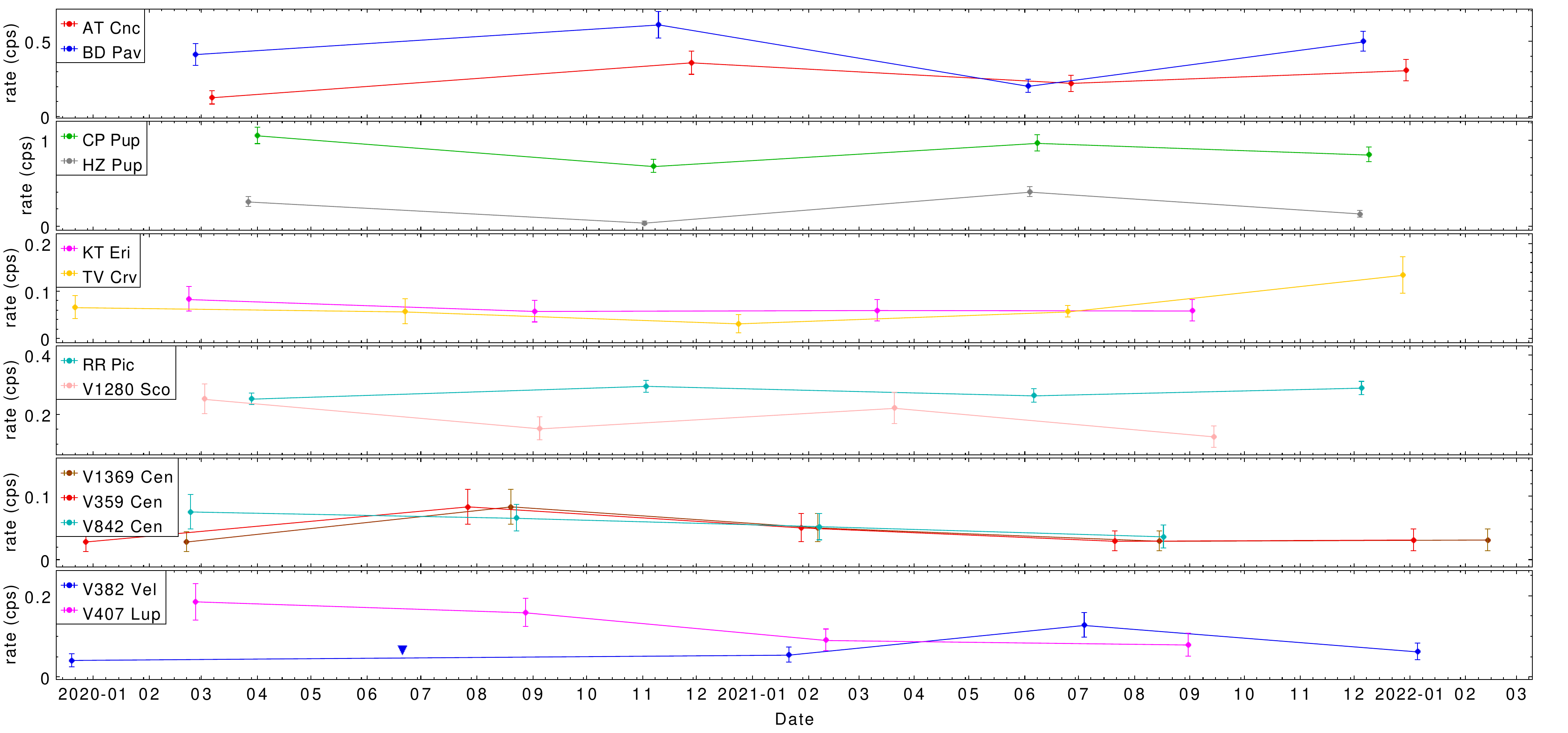}\hfill
       \caption{Variability in count-rate for novae detected in four or five eRASS. 
       }
    \label{lcs}
\end{figure*}

For the novae in the western Galactic hemisphere (eROSITA\_DE catalogue), we provide the detailed results for each individual survey as well as for the merged surveys in Table~\ref{tab:detailed_summary}. The count rate and total counts, uncertainty in the X-ray position and separation of the X-ray source from the SIMBAD coordinates of the optical nova, hardness ratios, and observation dates for each individual eRASS are reported in Table~\ref{tab:detailed_summary}. For some cases, the source is not detected in one or more of the individual eRASS, or even only detected in the merged eRASS:4 but not in any of the individual visits. We have checked in all cases that the upper limits for each individual eRASS with a non-detection of a source detected in the merged or other individual eRASS. We find that the upper limits on individual eRASS of sources detected in other eRASS are well compatible with the detected fluxes, and therefore not indicative of intrinsic X-ray variations of the source. Figure~\ref{lcs} shows the light-curves of the novae detected in at least four of the individual eRASS. 

\begin{figure}
    \includegraphics[width=0.47\textwidth]{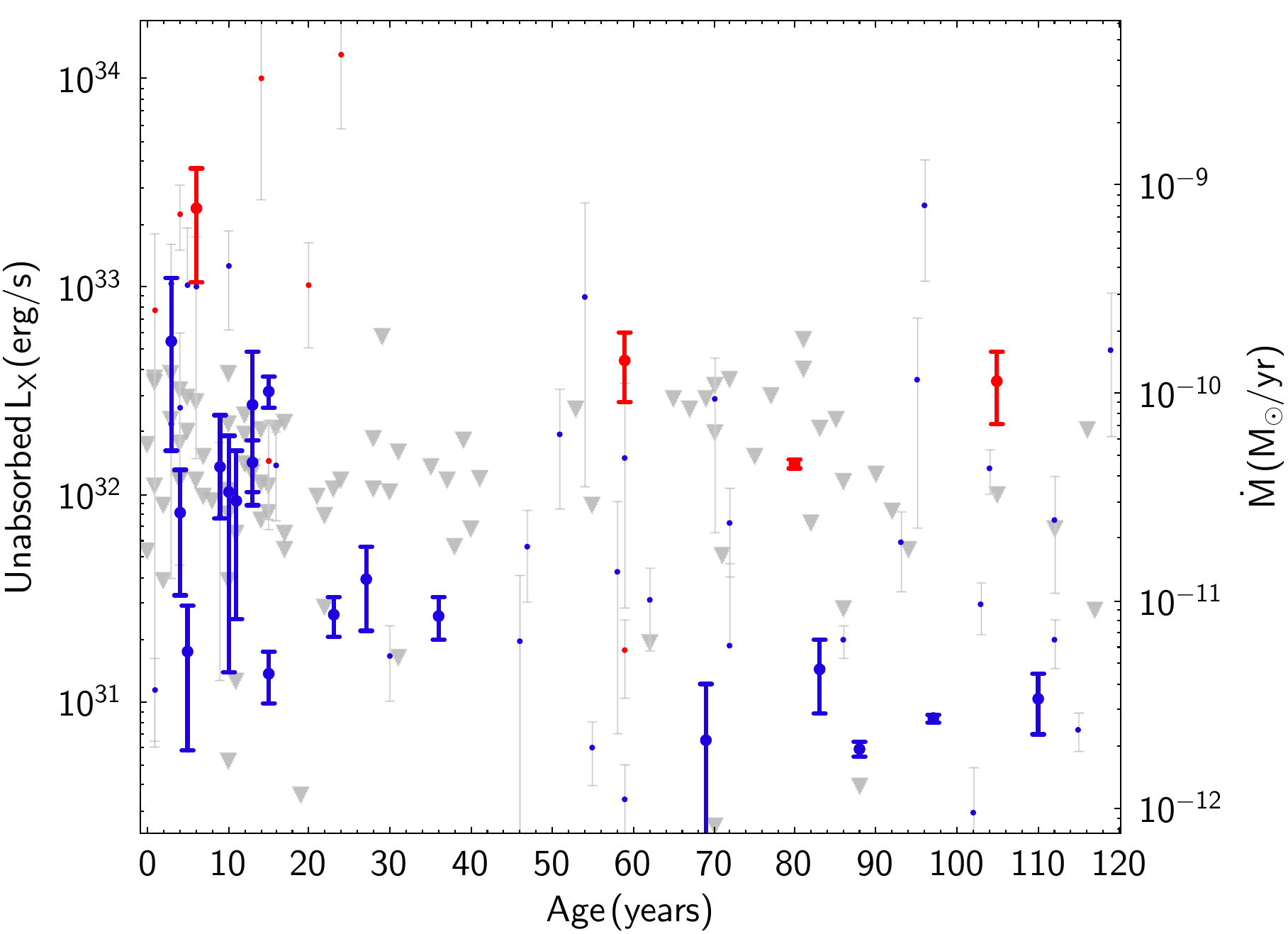}
       \caption{Unabsorbed X-ray luminosity as detected by eROSITA for historical novae located in the western Galactic hemisphere by eROSITA\_DE eRASS:4 (big points in red and blue with coloured error bars)  as a function of years since last nova outburst. Known magnetic systems are shown in red. Small points with gray error bars show the novae reported from the eastern Galactic hemisphere by \cite{2021AstL...47..587G}. Upper limits for both hemispheres are shown as grey triangles. }
    \label{mar_vs_age}
\end{figure}

With the distance and interstellar hydrogen column known for the sources in Table~\ref{tab:detailed_summary}, we can determine their unabsorbed X-ray luminosity. Since all these are cataclysmic variables years after the nova thermonuclear outburst, we can safely assume that accretion is the only source of X-ray emission and use the unabsorbed X-ray luminosity to estimate the accretion rate. Thus, we can provide for the first time a population study of the accretion rate of novae for the first decades after the outburst, shown in Fig.~\ref{mar_vs_age}. Since the cooling flow model (\mkcflow in XSPEC) is the model that best represents our available spectral data (see next section), we have obtained a conversion factor from the 0.2$-$2.3~keV unabsorbed luminosity to mass accretion rate (as part of the normalization of the \mkcflow model) by simulating eROSITA spectra in XSPEC for the range of luminosities of interest. With the conversion factor obtained, we show in Fig.~\ref{mar_vs_age} the evolution of accretion rate in the first 12 decades after a nova outburst. Known magnetic CVs are shown in red.

One interesting characteristic of the host system is the magnetic field of the white dwarf and, in particular, whether it is strong enough to disrupt the accretion disk and channel the accretion to the magnetic poles, i.e., whether the host system is a non-magnetic CV or an intermediate polar or polar. A key characteristic of the magnetic system is the reflection of the X-ray emission from the accretion column on the poles by the white dwarf surface: the heated surface emits as a blackbody, adding an extra soft component to the otherwise pure thermal plasma X-ray spectrum. Our sample has just a few sources with enough statistics for spectral analysis, so the search for such a soft component can be performed only for a few (see next section). However, for the low-statistic sources, a soft excess would in principle make the global X-ray spectrum softer and this could be identified in the hardness ratio, as shown for the global CV population in the eROSITA survey by \cite{2024A&A...690A.243S}. 

\begin{figure}
    \includegraphics[width=0.5\textwidth]{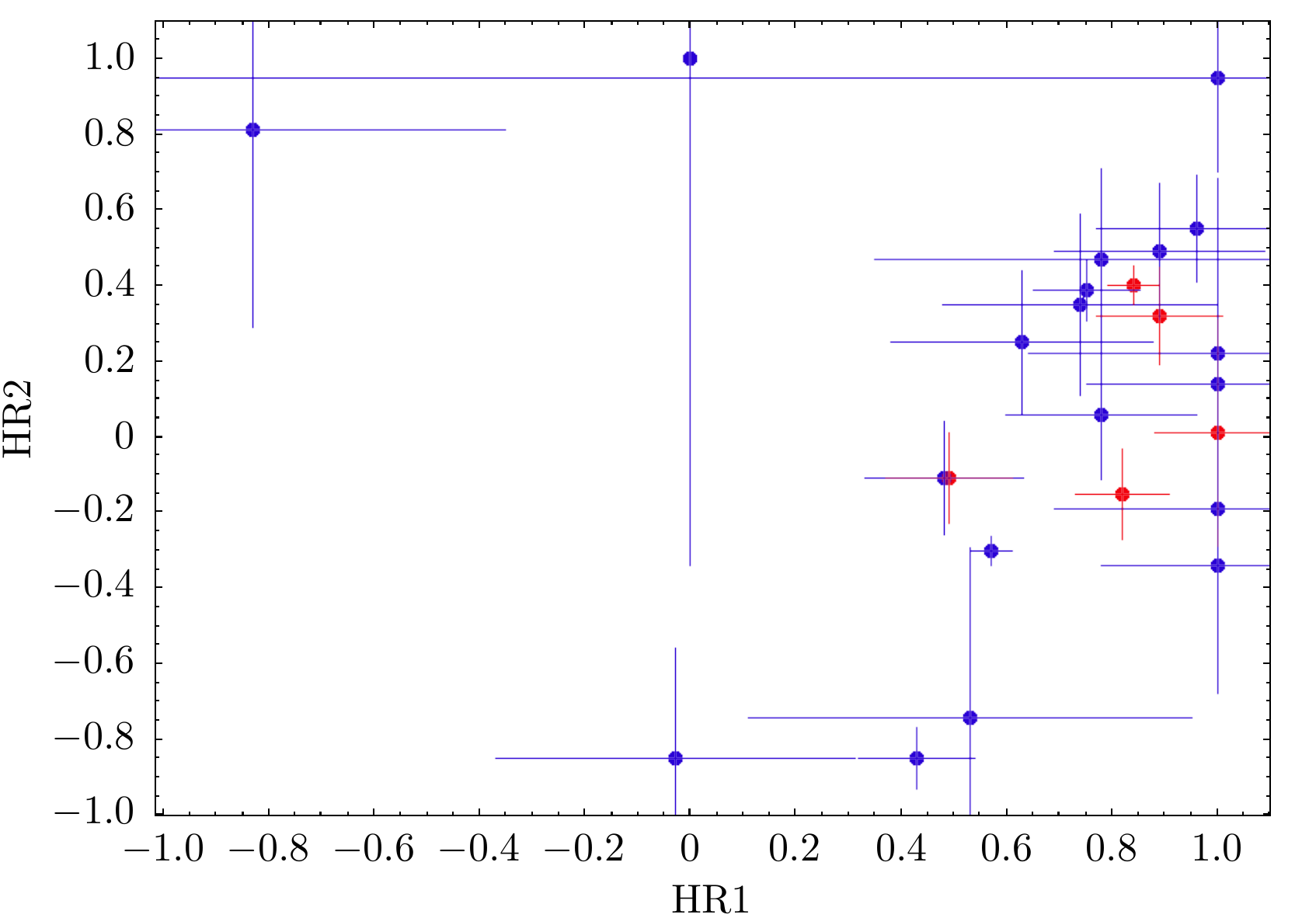}\hfill
       \caption{Hardness ratio plot for all eRASS:4 DE historical novae. Known magnetic systems are shown in red.}
    \label{HRplot}
\end{figure}

With the aim of identifying new magnetic candidates, we show the HR1-HR2 plot for our nova sample in Fig.~\ref{HRplot}, and list all data in Table~\ref{tab:detailed_summary}. We show the known magnetic systems in red, and find that they are not systematically softer in the hardness-ratio plot. As we will see for the particular case of AT Cnc in the next section, absorption plays a key role here, making the HR1-HR2 unsensitive to the presence of the soft component. 

\section{Spectral results: new bright X-ray counterparts to old novae}
\label{section_new_sources}

\subsection{AT~Cnc: an Intermediate Polar masquerading as a dwarf nova?}

AT~Cnc was discovered as a variable star in 1968 \citep{1968MmSAI..39..429R} and traditionally classified as a dwarf nova of the Z Cam type, showing normal outbursts and standstills typical of its class. The long-term light-curve available in the AAVSO archives since its discovery shows that standstills of AT~Cnc are unusually short (a few weeks), but sometimes it reveals abnormal very long standstills of about half a year, with visual magnitudes in the range of 13$-$14 (while the whole range of variation of its visual magnitude is 12.1$-$15.4). \citet{1999PASJ...51..115N} reported time-resolved spectroscopy during a standstill, measuring an orbital period of 0.2011$\pm$0.0006 days from the H$\alpha$ line. They also detected Na I absorption lines in the spectrum. The origin of both Na I and H$\alpha$ lines was unclear, disk or secondary, although this second option seems unlikely according to \citet{1997MNRAS.287..271S} because of the absence of TiO bands.

\citet{2012ApJ...758..121S} discovered a nova shell around AT~Cnc, indicating an old nova explosion 330 years ago \citep{2017MNRAS.465..739S}, which could correspond to the "guest star" reported in the constellation of Cancer by Korean observers in the year 1645~CE. They also determined the distance to AT Cnc to be 460 pc, consistent with results from Gaia DR3 ($455\pm7$\,pc, \citealt{2023AJ....165..163C}). 

\begin{figure}
    \includegraphics[width=0.5\textwidth]{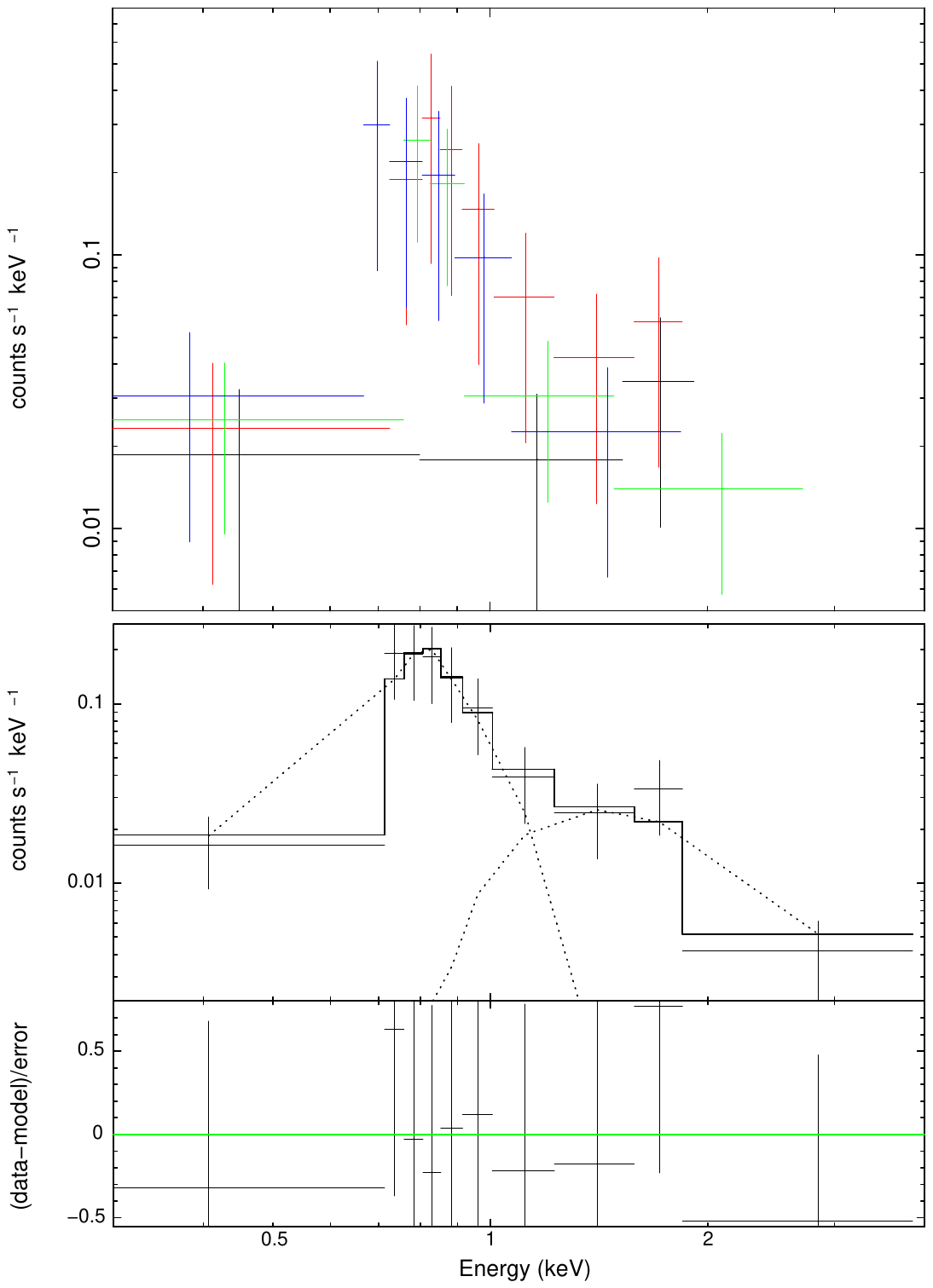}\hfill
    \caption{{\bf Upper panel}: AT Cnc spectra for the merged eRASS:4 (black) and for the individual eRASS1 (red), eRASS2 (blue), eRASS3 (green) and eRASS4 (magenta). {\bf Lower panel}: Merged spectrum for eRASS:4 with best fit model. Individual components of the model (blackbody at low energies and \mkcflow at higher energies) are shown in dashed lines.}
    \label{ATCnc_spec}
\end{figure}

\citet{2004A&A...419.1035K} reported the detection of superhumps, for the first time in a Z Cam system, indicating an asymmetric accretion disc. More recently the first suggestion of the possible magnetic nature of AT~Cnc appeared in the coherent brightness modulations reported by \citet{2019NewA...67...22B}: in addition to the orbital period of 0.2 days (290 min), they found a new modulation of 25.7 minutes, most probably corresponding to the white dwarf spin. As the authors suggested, all this points to the possibility of an intermediate polar scenario.

 eROSITA provides the first registered X-ray detection of the system.  Individual eRASS spectra are shown in Fig.~\ref{ATCnc_spec} and lack the S/N for spectral analysis. It is, however, evident that in the first eRASS the count rate is much lower than in the following three, with just a marginal detection in eRASS1. The spectral energy distribution remains similar for the three following spectra obtained 0.5, 1 and 1.5 years afterwards. We consider the merged eRASS:4 spectra for spectral analysis. The total count rate from eRASS:4 extracted from the cataloge is 0.254$\pm$0.031 \cps. For spectral fitting, we use the merged eRASS:4, manually extracted spectrum, with a binning containing at least 5 counts per bin, which once in XSPEC and after background subtraction has a net count rate of 0.09$\pm$0.02 cts/s, and a total exposure time of 571\,s.  

At first glance, the eRASS:4 spectrum of AT Cnc does not seem to correspond to a single component model (see Fig.~\ref{ATCnc_spec}). The detected rate is maximum at 0.8~keV,
and the count rate below 0.7~keV is negligible. Taking the results with caution due to the low statistics, the observed spectrum is well reproduced by a soft bbody component plus a bremsstrahlung, or by a more realistic isobaric cooling flow thermal plasma (\mkcflow model), with a low temperature of 500 eV and maximum temperature of 3~keV, but with unconstraining uncertainties. For the other parameters, the uncertainties are determined with the steppar function and contour plots for the parameters of the black body and the plasma model separately (see Fig.~\ref{ATCnc_cont}), freezing the parameters of the other model component to best-fit values. The accretion rate is $5(\pm3)\times10^{-11}$\,M$_{\odot}$ yr$^{-1}$. The soft excess is well fit with a blackbody with kT=$50\pm10$~eV and a lower limit for the bolometric luminosity of $10^{34}$\,\lx. The 0.2$-$2.3~keV flux with this best fit model to the merged eRASS:4 spectrum is $3.0(\pm0.5)\times10^{-13}$\,\fergs,  corresponding to a total 0.2$-$2.3~keV luminosity $7(\pm1)\times10^{30}$\,erg s$^{-1}$. We have used the Upper Limit ESA server HILIGT\footnote{http://xmmuls.esac.esa.int/upperlimitserver/} to check that the upper limits for the 0.2$-$2.3~keV flux from the ROSAT All Sky Survey (<8$\times10^{-13}$\,\fergs) and the \xmm Slew Catalogue (<4$\times10^{-13}$\,\fergs) are compatible with the flux detected by eROSITA.

The position of AT Cnc in the hardness ratio is surprisingly hard for a source with a soft component (HR12$\sim$0.8, see Table \ref{tab:detailed_summary}). The reason is that the count rate below 0.7~keV is negligible, suggesting a source of intrinsic absorption. A spectral fit including the lowest energy spectral channels yields an absorbing column of $2(\pm1)\times10^{22}$\,cm$^{-2}$. The average interstellar hydrogen column in the direction of AT Cnc is $3.8\times10^{20}$cm$^{-2}$ from the H14PI maps \citep{2016A&A...594A.116H}. Since AT Cnc is at a larger distance to the Sun than the sources included in the H14PI maps in the direction of interest, we check the reddening from the STILISM maps, which indicate E(B-V)$=0.3\pm0.1$ \citep{2014A&A...561A..91L, 2017A&A...606A..65C}. This corresponds to an hydrogen column of $1.6(\pm0.1)\times10^{20}$ cm$^{-2}$ (using the most widespread E(B-V) relation from \citealt{1977ApJ...216..291S} and \citealt{1978ApJ...224..132B}) or $2.3(\pm0.2)\times10^{20}$ cm$^{-2}$ with a more recent determination for the N$_H$/E(B-V) relation \citep{2014ApJ...780...10L}. So a local or intrinsic source of absorption seems to be present. \citet{2012ApJ...758..121S} determined a total mass in the nova shell of $5\times10^{-5}M_\odot$ and a shell radius of 0.2 pc. Assuming an homogeneous distribution of the ejected mass and pure atomic hydrogen, the column density in the line of sight due to the shell would be only $10^{17}$ cm$^{-2}$. However, the shell is clearly not homogeneous and the images show that the matter is in blobs of circa 1\arcsec\ in diameter \citep{2017MNRAS.465..739S}.

\subsection{BD Pav}

BD Pav is a dwarf nova of the U Gem type located at 329$\pm$2 pc, which underwent a nova eruption in 1934. \citet{1983A&A...124..287B} determined the orbital period to be 4.3 hours, with an exceptionally bright secondary. The light curves in quiescence showed sharp eclipses, suggesting a relatively high inclination system. \citet{2008ApJ...681..543S} attempted to determine the white dwarf mass from HST STIS spectra (they fit the data with a 1.2~M$_{\odot}$ white dwarf with an effective temperature of 27000~K and an inclination of 75º for the optically thick disk), but the fit was not good, and the results are to be taken with care. Interstellar hydrogen column in the direction of BD Pav is 6$\times10^{20}$cm$^{-2}$ \citep{2016A&A...594A.116H}. 

\begin{figure}
    \includegraphics[width=0.5\textwidth]{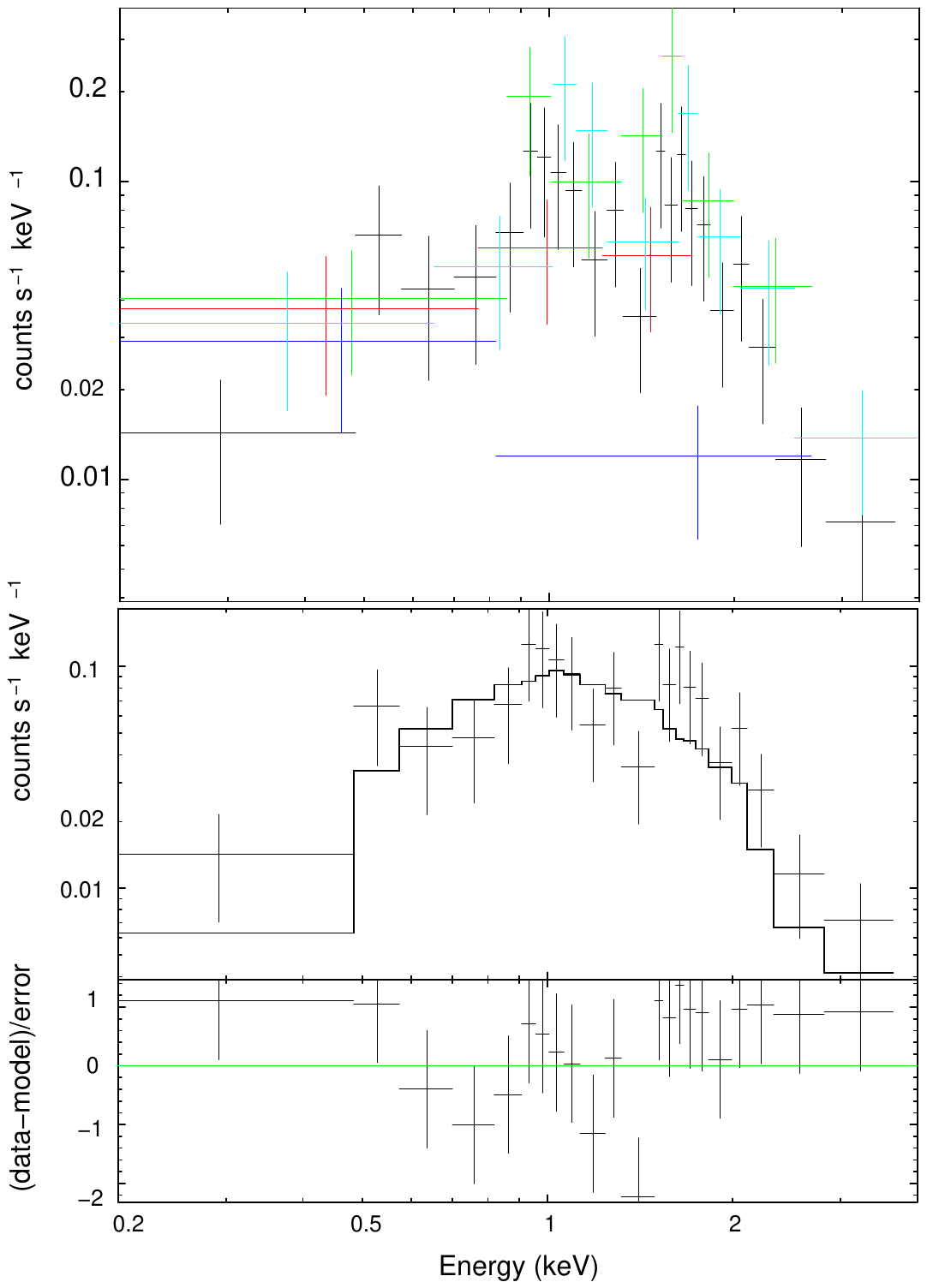}\hfill
\caption{{\bf Upper panel}: BD Pav spectra for the merged eRASS:4 (black) and for the individual eRASS1 (red), eRASS2 (blue), eRASS3 (green) and eRASS4 (magenta). {\bf Lower panel}: Merged spectrum for eRASS:4 with best fit model. 
\label{BDPav_spec}}
\end{figure}

Individual eRASS have very low S/N (see Fig.~\ref{BDPav_spec}), so we work with the merged eRASS:4 results. The total count rate from eRASS:4 extracted from the catalogue is 0.423$\pm$0.034 \cps. For spectral fitting, we use the manually extracted eRASS:4 spectra, with a binning containing at least 5 counts per bin (Fig.~\ref{BDPav_spec}). The spectral data used to fit have a net count rate of 0.13$\pm$0.02 \cps, with a total exposure time of 811\,s. A cooling flow \mkcflow model can reproduce the spectrum with a high temperature larger than 40~keV (1$\sigma$ lower limit; upper limit unconstrained, see Fig.~\ref{BDPav_spec_cont}), an accretion rate of 5($^{+7}_{-2})\times10^{-14}$\,M$_{\odot}$ yr$^{-1}$ and an absorbing hydrogen column density of $2(_{-1}^{+2})\times10^{21}$cm$^{-2}$.  A simple absorbed bremsstrahlung provides a similar fit but with unconstrained temperature. The 0.2$-$2.3~keV flux with the best fit \mkcflow model is 5(${\pm1})\times10^{-13}$\,\fergs, corresponding to a total 0.2$-$2.3~keV luminosity 6.5$(\pm 2.0)\times10^{30}$\,\lx.
The flux is compatible with the upper limit from the \xmm Slew catalogue. There are no published results for the X-ray source and the only previous X-ray detection found in the archives is that of the ROSAT All Sky Survey, in August 1990, with a detection of the source with a count rate of 0.079$\pm$0.030 \cps, corresponding to a flux of 7.5(${\pm3.2})\times10^{-13}$\,\fergs (0.2$-$2.3~keV), also compatible with the flux determined now by eROSITA.\footnote{For the ROSAT flux calculation, we first fit a power-law to the eRASS:4 spectrum to find the best fit absorption column and slope, and check that the total flux for the eROSITA merged spectrum is the same as for the best fit model. Then we use this power-law parameters for conversion of the ROSAT count rate to flux.} 

\section{Nova hosts CVs previously known in X-rays: spectral analysis}
\label{section_old_sources}

\subsection{CP Pup}

CP Pup is a well-known CV, with a nova outburst in 1942. 
\cite{2009ApJ...690.1753O} obtained \xmm/RGS grating spectra, and their results suggested a high mass and magnetic white dwarf. The \xmm spectra obtained in 2005 were well fit with an isobaric cooling model, \mkcflow, with a maximum temperature of 80~keV and an absorbing hydrogen column of $2\times10^{21}$cm$^{-2}$, with a flux in the RGS (0.33$-$2.5~keV) energy range of $2.3(\pm0.2)\times10^{-12}$\,\fergs. The normalization constant of the cooling flow model indicated an upper limit for the accretion rate of $8\times10^{-11}$\,M$_{\odot}$ yr$^{-1}$ for a distance of 850~pc, the minimum distance estimation at the time. 
\cite{2013MNRAS.436..212M} based on \chandra/HETG refined the accretion rate to $3.3\times10^{-10}$\,M$_{\odot}$ yr$^{-1}$ with a maximum temperature of the cooling model of 40($\pm20$)~keV and find that the X-ray spectral characteristics are consistent with those of known IPs.

\begin{figure}
    \includegraphics[width=0.5\textwidth]{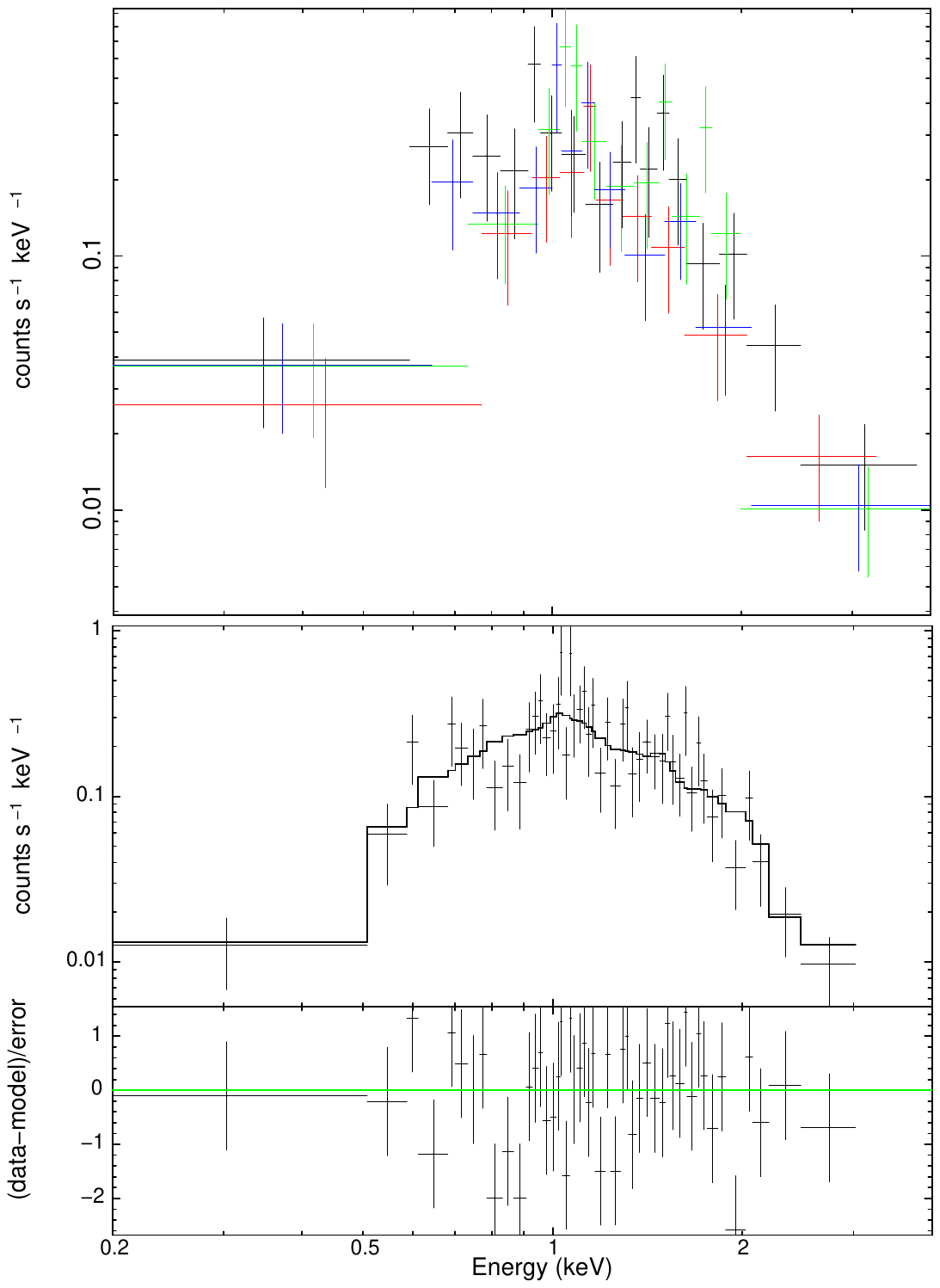}\hfill
   \caption{\label{CPPup_spec} Same as Fig.~\ref{BDPav_spec} for CP Pup}
\end{figure}

 The total count rate from eRASS:4 extracted from the catalogue is 0.89$\pm$0.04 \cps. For spectral fitting, we use the manually extracted eRASS:4 spectra, with a binning containing at least 5 counts per bin (Fig.~\ref{CPPup_spec}). A cooling flow \mkcflow model can reproduce the spectrum with a high temperature of 15($^{+40}_{-7})$\,keV  and an accretion rate of 5($^{+5}_{-3})\times10^{-11}$\,M$_{\odot}$ yr$^{-1}$, with $N_H=3.0(\pm1.5)\times10^{21}$\,cm$^{-2}$. The residuals show a possible absorption around 0.65~keV (possibly O VIII). The 0.2$-$2.3~keV flux with the best fit \mkcflow model is 9$(\pm5)\times10^{-13}$\,\fergs. This flux corresponds to a total 0.2$-$2.3~keV luminosity 6$(\pm 3)\times10^{31}$\,\lx 
 \citep[at 767 pc as determined from Gaia DR3,][]{2023AJ....165..163C}.

\subsection{HZ Pup}

Evidence of the magnetic nature of HZ Pup (with a 5.11 h orbital period) comes from the identification of a possible spin period \citep[20 minutes,][]{1997ASPC..121..679A} also detected in the \xmm light curve by \cite{2020A&A...639A..17W}. The same authors fit the faint-phase spectra with a single temperature APEC (15 keV) and add a second APEC model for the bright phase (64 keV). They also add two Gaussian emission lines at 0.5 keV, to account for the soft excess, and at 6.4 keV for the fluorescence Fe line. In Table 4 of \cite{2020A&A...639A..17W} it is indicated that the X-ray flux for each of the two APEC components used to fit the \xmm spectra is $6\times10^{-12}$\,\fergs. The distance to HZ Pup is 2.7($\pm0.5)$~kpc from Gaia DR3 \citep{2023AJ....165..163C}.

\begin{figure}
    \includegraphics[width=0.5\textwidth]{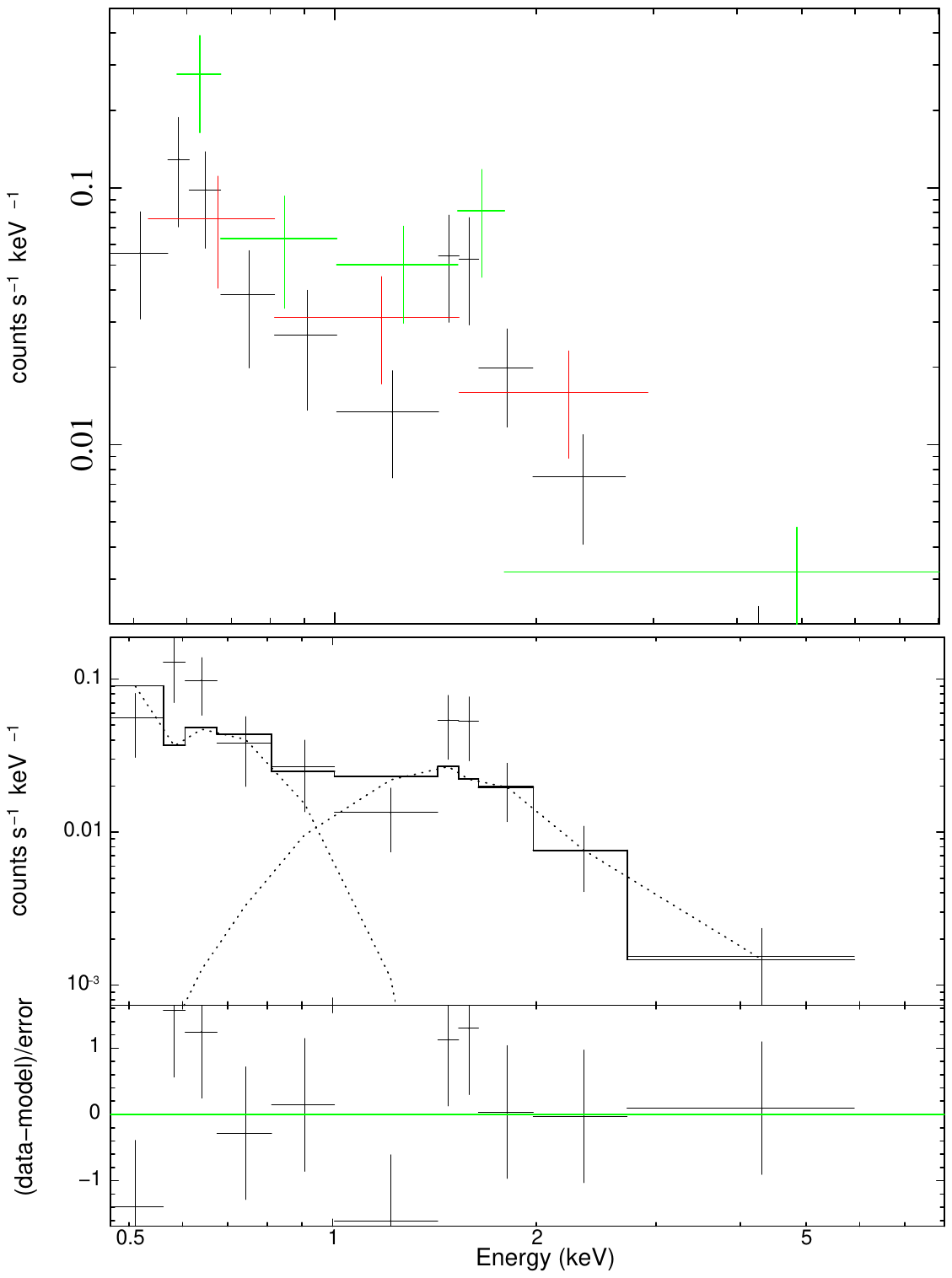}\hfill
    \caption{\label{HZPup_spec}  Same as Fig.~\ref{BDPav_spec} for HZ Pup}
\end{figure}

 The total count rate from eRASS:4 extracted from the catalogue is 0.227$\pm$0.024 \cps. For spectral fitting, we use the merged eRASS:4 manually extracted spectra, with a binning containing at least 5 counts per bin (Fig.~\ref{HZPup_spec}). 
 
 A single cooling flow \mkcflow model fails to reproduce the spectrum, which shows a clear soft excess. Even allowing an absorbing hydrogen column orders of magnitude lower than from the ISM, the plasma model is too flat to reproduce the soft excess. Thus we add a blackbody model to fit the soft excess. The isobaric cooling flow has a low temperature 0.1$-$0.2 keV (free fits keep around this value even without blackbody component and without frozen parameters), while the maximum temperature pegs at a maximum of 80 keV (both without and with the blackbody components present). The accretion rate for the best-fit model is 6.7$\times10^{-11}$\,M$_{\odot}$ yr$^{-1}$. The blackbody accounts for the soft excess with a temperature of 50 eV and a normalization of 4$\times10^{-4}$ which corresponds to $3\times10^{34}$ erg s$^{-1}$. The X-ray flux 0.2$-$2.3 keV is $2\times10^{-13}$\,\fergs (corresponding X-ray luminosity at 2.7 kpc is $2\times10^{32}$ erg s$^{-1}$).

\subsection{RR Pic}

RR Pic is classified in the Ritter \& Kolb catalogue as an SW Sex star; Orio et al. (2001) considered it to be the only suspected magnetic CV in the old nova ROSAT sample. However, the most recent \chandra observation (Pekön \& Balman 2008) found no evidence of a magnetic nature, with the ACIS-S spectra fit with a vmcflow model, with some overabundance of C and O, and a maximum temperature of around 2 keV. The distance to RR Pic is now known to be 492($\pm5)$~pc from Gaia DR3 
\citep{2023AJ....165..163C}.

\begin{figure}
    \includegraphics[width=0.5\textwidth]{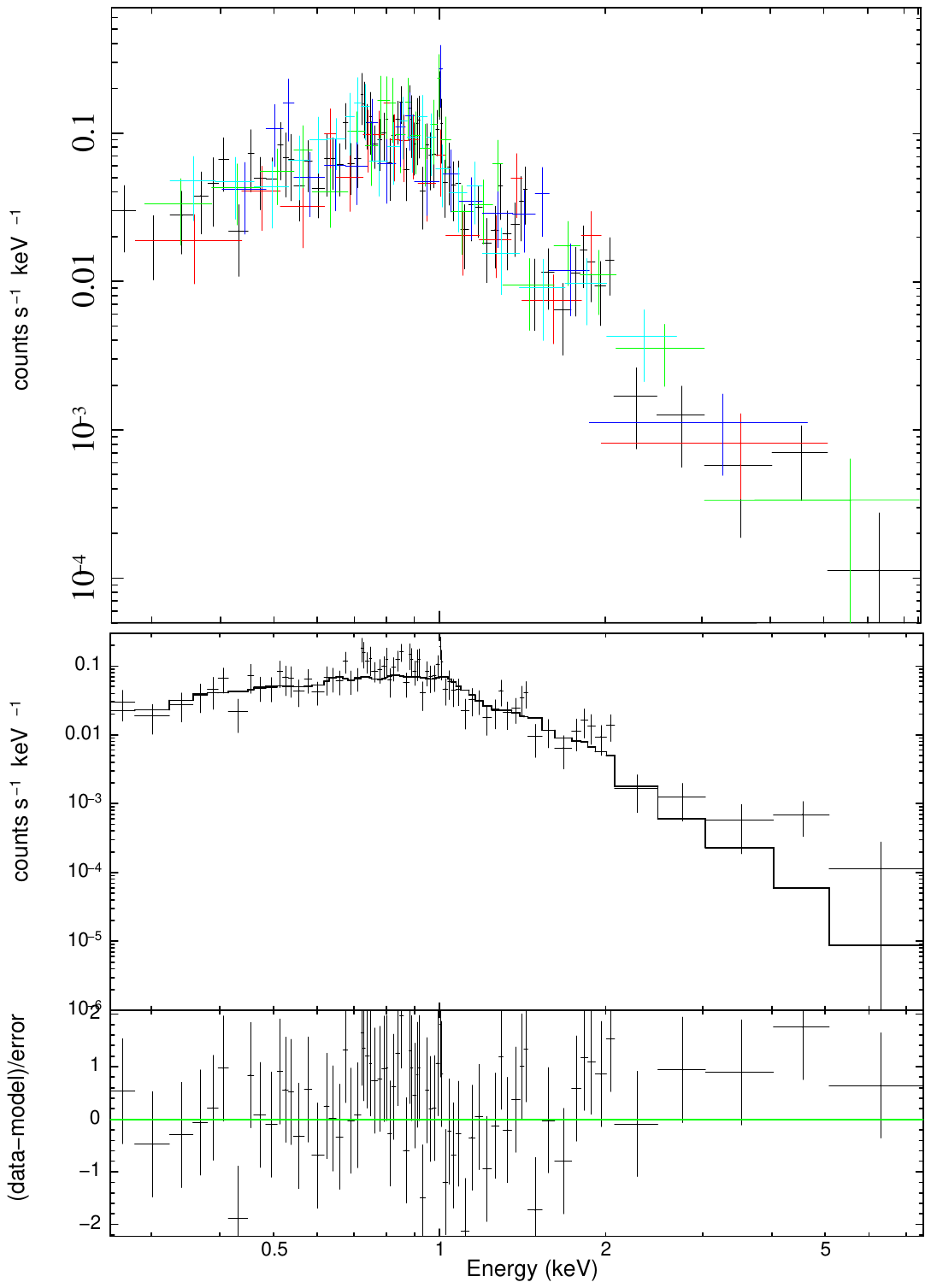}\hfill
    \caption{\label{RRPic_spec} Same as Fig.~\ref{BDPav_spec} for RR Pic}
\end{figure}

 The total count rate from eRASS:4 extracted from the catalogue is 0.280$\pm$0.010 \cps. For spectral fitting, we use the merged eRASS:4, manually extracted spectra, with a binning containing at least 5 counts per bin (Fig.~\ref{RRPic_spec}). The net rate of the spectrum used to fit is 0.073$\pm$0.004 \cps, with a total exposure time of 5904\,s (448 spectral counts). A cooling flow \mkcflow model reproduces the spectrum leaving some excess above 2 keV (as found by Pekön \& Balman 2008 in \chandra spectra). The isobaric cooling flow model that best reproduces the data has a low temperature of 80$(\pm20)$~eV and a maximum temperature of 3.5$(^{+1.5}_{-0.8})$~keV, with an accretion rate 14$(^{+7}_{-5})\times10^{-12}$\,M$_{\odot}$ yr$^{-1}$.
 The X-ray flux 0.2$-$2.3 keV is $2.5\times10^{-13}$\,\fergs (corresponding X-ray luminosity at the distance of RR Pic is $7\times10^{30}$ erg s$^{-1}$).

\subsection{A not-so-old nova, now in quiescence: V407 Lup 2016}

V407 Lup (Nova Lup 2016) is thought to have occurred on an intermediate polar. \citet{2018MNRAS.480..572A} studied the X-ray SSS emission and found evidence of resumed accretion. They found two periodicities in UV and X-rays at 3.57\,h and 565\,s, interpreted as the orbital period and the spin of the white dwarf in an intermediate polar, which was confirmed with \xmm observations obtained in quiescence in 2020 \citep{2024MNRAS.533.1541O}. \citet{2018MNRAS.480..572A} also report a complex X-ray spectrum, including a hard component which could be fit by a flat power law or a relatively high-temperature thermal component. 

The eRASS:4 spectrum of the quiescent nova is well fit with a thermal plasma. A single high-temperature Mekal (80 keV) fits reasonably well, similarly to a \mkcflow model with low temperature 100 eV, maximum temperature pegged at 80 keV, and an accretion rate of 3$\times 10^{-11}$\,M$_{\odot}$ yr$^{-1}$ (for a 3 kpc distance). With the large uncertainty in the distance \citep{2018MNRAS.480..572A}, the statistical uncertainties are not significant. Still, the spectral parameters are consistent with an accreting, isobaric shock flow on a white dwarf.  The X-ray flux 0.2$-$2.3 keV is $1.2\times10^{-13}$\,\fergs (corresponding X-ray luminosity $10^{32}$ erg s$^{-1}$).

\begin{figure}
    \includegraphics[width=0.5\textwidth]{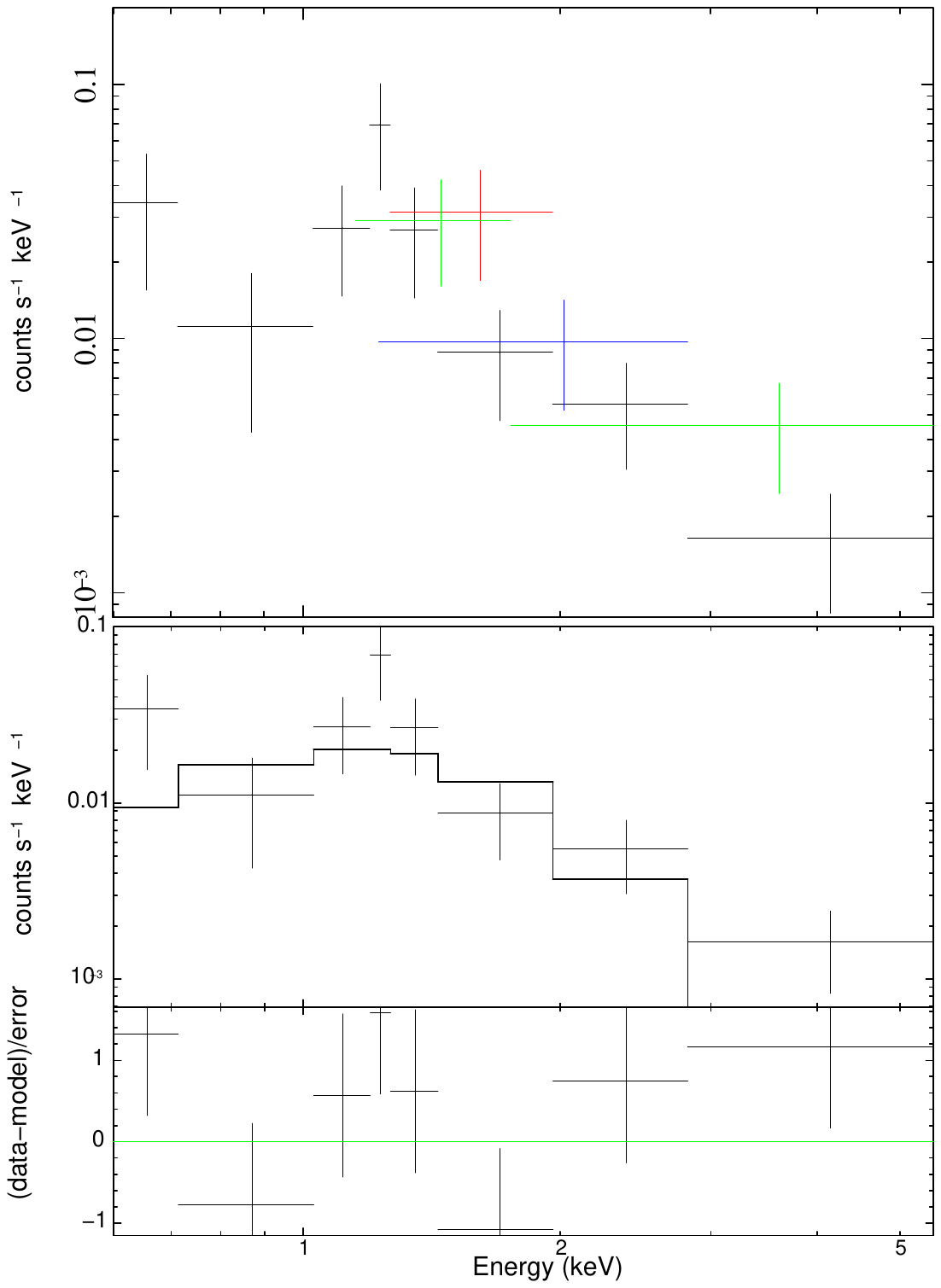}\hfill
    \caption{\label{V407_spec} Same as Fig.~\ref{BDPav_spec} for V407 Lup}
\end{figure}

\begin{table*}
\centering
\caption{Best-fit spectral parameters of novae with more than 50 \cps in the merged spectra. Uncertainties are given at approx. 90\%, but are better constrained in the confidence contours in the associated figures for AT~Cnc and BD~Pav.}
\label{tab:spec_summary}
\begin{tabular}{lccccc}
\hline\noalign{\smallskip}
Nova & $N^{\mathrm{TBabs}}_H$ & $kT_{\mathrm{bbdy}}$ & $kT^{\mathrm{max}}_{\mathrm{mkcflow}}$ & $\dot{M}$ & $L_{\mathrm{X-ray}}$ (0.2--2.3 keV) \\
& ($10^{21}$ cm$^{-2}$) & (eV) & (keV) & $(10^{-11} M_{\odot}$ yr$^{-1}$) & ($10^{31}$ erg s$^{-1}$) \\
\noalign{\smallskip}\hline \hline\noalign{\smallskip}
AT Cnc & $20 \pm 10$ & $50 \pm 10$ & $\sim 3.0$ & $5 \pm 3$ & $0.7 \pm 0.1$ \\
BD Pav & $2^{+2}_{-1}$ & -- & $>40$ & $0.005^{+0.007}_{-0.002}$ & $0.65 \pm 0.2$ \\
CP Pup & $3.0 \pm 1.5$ & -- & $15^{+40}_{-7}$ & $5^{+5}_{-3}$ & $6 \pm 3$ \\
HZ Pup & $10^{+15}_{-5}$ & $100 \pm 50$ & $>80$ & $2^{+3}_{-1}$ & $20 \pm 10$ \\
RR Pic & $0.2^{+0.3}_{-0.1}$ & -- & $1.7^{+0.7}_{-0.5}$ & $1.4^{+0.7}_{-0.5}$ & $0.7^{+0.5}_{-0.2}$ \\
V407 Lup & $2$ (unconst) & -- & $80$ (unconst) & $3$ (unconst) & $10$ (unconst) \\
\hline
\end{tabular}
\footnote{Uncertainties are given at approx. 90\%, but are better constrained in the confidence contours in the associated figures for AT~Cnc and BD~Pav}
\end{table*}

\section{Discussion}
\label{section_discussion}

The evolution of accretion rate in the first 12 decades after a nova outburst shown in Fig.~\ref{mar_vs_age} clearly indicates a lack of bright sources after the first two decades. The right vertical axis shows the accretion rate corresponding to the observed unabsorbed flux, determined as explained in Sect.~\ref{section_population}. Known magnetic CVs are shown in red, and clearly show systematic higher X-ray luminosities, and therefore, higher X-ray luminosities than non-magnetic systems. While our procedure may have introduced some systematics in the values of the accretion rates obtained, since the factor is constant for all cases, the evolution as a function of age is not affected by the particular value of the conversion from X-ray luminosity to accretion rate. Therefore, the result of the accretion rate evolution after the nova outburst is robust: there is a clear excess of high-accretion rate systems shortly after the nova outburst, decreasing during the first two decades. After that initial decrease, the accretion rates seem to remain at a similar level for the first century after the outburst. A second view of the same effect free of any assumptions is found simply in the fraction of novae detected in X-rays as a function of time after outburst, in Fig.~\ref{histogram_all_x_frac}.  

One aspect of concern to be considered when assessing the accretion-powered X-ray emission of faint CVs is the possible contribution from the corona of the donor star to the total X-ray luminosity. \citet{2004A&A...417..651S} showed from nearby stars in the ROSAT All Sky Survey that the X-ray luminosity of F/G/K/M stars very rarely reaches values as large as $10^{29}$ \lx, with most stars emitting at $10^{27}-10^{28}$ \lx. This is well below the faintest sources of our sample and thus we can assume that the X-ray luminosities observed are accretion powered. 

The evolution suggested by Fig.~\ref{mar_vs_age} agrees with the predicted behaviour of accretion in post-outburst nova system as modeled by \citet{2020NatAs...4..886H}. The enhanced accretion rate shortly after the nova outburst is explained by the irradiation of the secondary by the hot white dwarf during and short after the nova eruption. This causes the expansion of the secondary, enhancing Roche Lobe overflow and the mass transfer, which ultimately causes an increased accretion rate on to the white dwarf and therefore increased accretion-powered X-ray luminosities. Accretion rate is then predicted to decrease by \citet{2020NatAs...4..886H} over a few hundred to thousands of years, as the effect of the white dwarf irradiation diminishes. We can provide here the first X-ray observational test for this predicted behaviour. From Fig.~\ref{mar_vs_age} we can confirm the initial decay of the accretion rate. This decay, however, seems to be predominant only for the first 30 years. Our observed X-ray luminosities of quiescent novae seem to stop decreasing after that, indicating that either the white dwarf cooling is faster, or the effect of the irradiated white dwarf on the secondary may be shorter than predicted. 

For spectral properties, we limit our studies to the western Galactic hemisphere. A point of interest is to identify new candidates to magnetic systems, and for that we plot the HR1-HR2 hardness ratio diagram in Fig.~\ref{HRplot}. The systematic study of cataclysmic variables observed by eROSITA presented by \citet{2024A&A...690A.243S} shows the distribution of non-magnetic, intermediate polar and polar CVs in the HR1-HR2 hardness plot. It is clear that non-magnetic systems cluster at a higher value of HR1, and only polars and some intermediate polars appear at lower values of HR1. But it is also clear that many of the magnetic systems show HR1-HR2 positions in the region of the non-magnetic systems. 

\begin{figure}
    \includegraphics[width=0.5\textwidth]{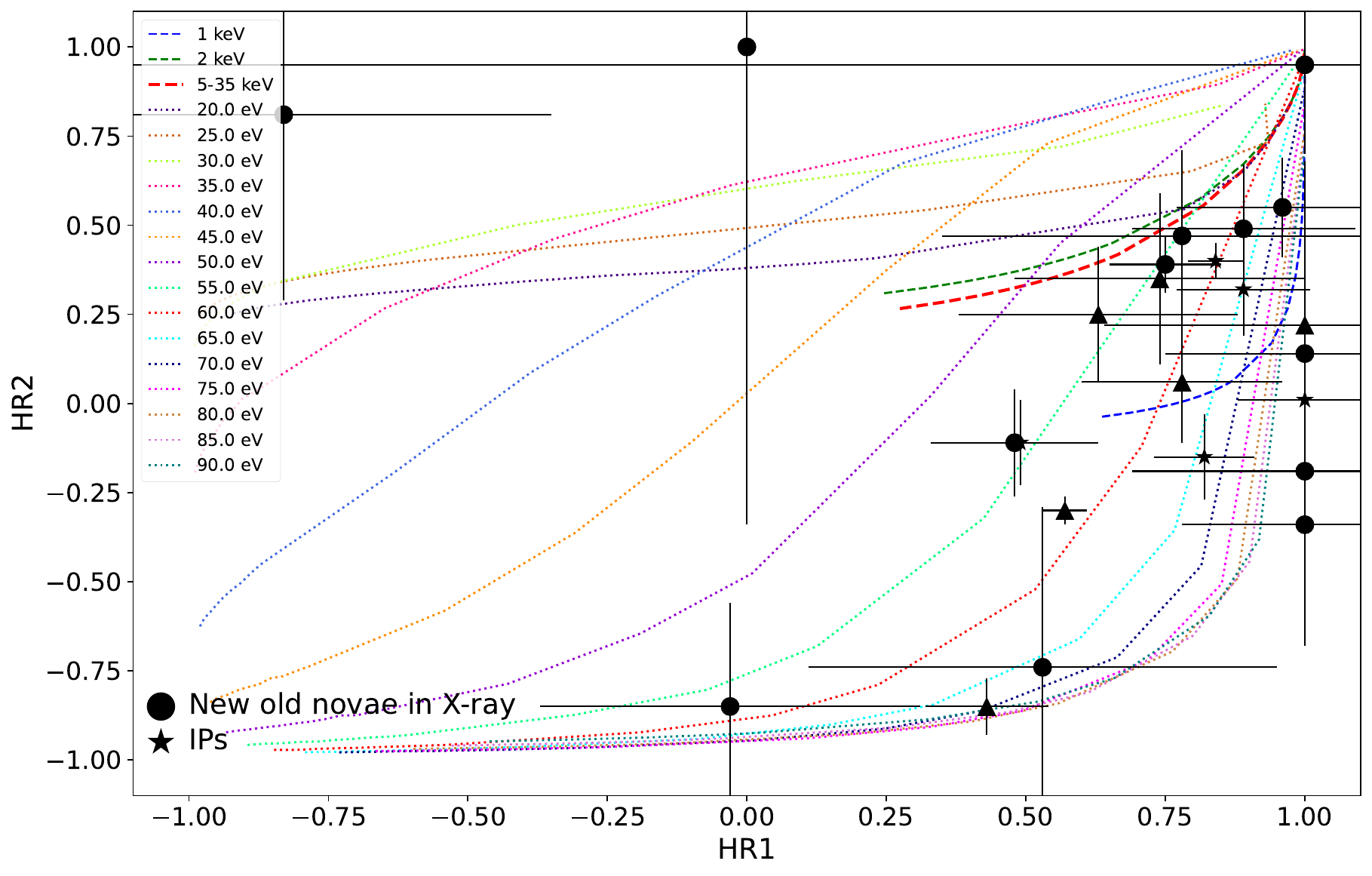}\hfill
       \caption{Hardness ratio plot with our detected novae and simulated paths for spectral models with variable absorption hydrogen column (increasing between $1\times10^{20}$ to $2\times10^{22}$ cm$^{-2}$ from left to right). Pure thermal plasma models (mekal) are shown in dashed lines, with kT increasing from 1 keV (blue, dashed line) to 35 keV. All models with kT above 5 keV collapse on the same path on the plot (red dashed line).Dotted lines show models with an extra soft, blackbody component, with effective temperatures from 20 eV (purple) to 90 eV (green). The simulations with soft component also include a thermal plasma with kT=15 keV and a normalization a factor 10 fainter than the blabkbody .  Detected historical novae in the eRASS:4 are shown as big dots for new detections, stars for known intermediate polars, and triangles for all other cases.}
    \label{HR_sim}
\end{figure}

From our HR1-HR2 plot (Fig.~\ref{HRplot}) we immediately find that most known magnetic systems do not cluster in a particular region of the hardness ratio plot. From the cases were we do have spectra, as is the case of AT Cnc, it seems clearly related to the role of absorption. To further explore the role of absorption in the position of magnetic systems in the HR1-HR2 plot, we have simulated the tracks on that plot for a set of spectral models in Fig.~\ref{HR_sim}. We plot the position of simulated eROSITA spectra in the HR1-HR2 diagram in Fig.~\ref{HR_sim} for a series of models only including thermal plasma component, or adding also a soft, blackbody component (indicative for the accretion on the magnetic poles). For each set of models, we simulate the HR1-HR2 tracks for increasing absorption column. It is clear that even with the presence of a soft component, absorption can place the source in the same region of the HR1-HR2 plot as non-magnetic systems.

\section{Summary}
\label{section_summary}
We have presented a large-scale X-ray population analysis of historical novae observed during the first two years of the eROSITA All Sky Survey (eRASS:4), focusing on novae in the western Galactic hemisphere. We have cross-matched 467 cleaned nova candidates with eROSITA source catalogues and identified 30 old novae in X-rays, 17 of which were detected in X-rays in quiescence for the first time. When combined with previous detections from the eastern hemisphere \cite{2021AstL...47..587G}, this yields an 18\% X-ray detection rate for historical novae in quiescence. The detection of novae of the XX and XXI century allows for showing a clear trend: accretion rates are higher in the first 20–30 years post-outburst, then stabilize over the next century. A hardness-ratio analysis was attempted to assess the magnetic nature of the white dwarfs, although absorption often obscures soft X-ray components. Some sources are bright enough for spectral analysis, and among them, AT Cnc was identified as a likely intermediate polar based on the clear soft excess of its X-ray spectrum.

\begin{acknowledgements}
This work is based on data from eROSITA, the soft X-ray instrument aboard \srg, a joint Russian-German science mission supported by the Russian Space Agency (Roskosmos), in the interests of the Russian Academy of Sciences represented by its Space Research Institute (IKI), and the Deutsches Zentrum f{\"u}r Luft- und Raumfahrt (DLR). The \srg spacecraft was built by Lavochkin Association (NPOL) and its subcontractors, and is operated by NPOL with support from the Max Planck Institute for Extraterrestrial Physics (MPE).
The development and construction of the eROSITA X-ray instrument was led by MPE, with contributions from the Dr. Karl Remeis Observatory Bamberg \& ECAP (FAU Erlangen-N{\"u}rnberg), the University of Hamburg Observatory, the Leibniz Institute for Astrophysics Potsdam (AIP), and the Institute for Astronomy and Astrophysics of the University of T{\"u}bingen, with the support of DLR and the Max Planck Society. The Argelander Institute for Astronomy of the University of Bonn and the Ludwig Maximilians Universit{\"a}t Munich also participated in the science preparation for eROSITA.
The eROSITA data shown here were processed using the \eSASS/\nrta software system developed by the German eROSITA consortium.
GS acknowledges support by the Spanish MINECO grant PID2023-148661NB-I00, by the E.U. FEDER funds, and by the AGAUR/Generalitat de Catalunya grant SGR-386/2021. GS acknowledges support from the Max-Planck-Institut f\"ur extraterrestrische Physik for her summer visits in 2023, 2024, and 2025.
\end{acknowledgements}

%
%
\bibliographystyle{aa} 
\bibliography{myrefs.bib} 
\onecolumn 
\begin{appendix}\section{Source lists}

\begin{center}
\begin{longtable}{ccccccc}
\caption{Historical novae correlated with eRASS:3, eRASS:4 or eRASS5 X-ray detections. IDs indicated as {\it sm} indicate merged survey from which the detected rate is reported, while {\it em} corresponds to a single eRASS detection, meaning the source was detected in only one of the surveys. Interstellar column densities are given for the coordinates and distance to each source using the 3DN$_H$-tool \citep{2024arXiv240303127D}.}
\label{tab:summary}
   \\ \hline\noalign{\smallskip}
      CV Name & Nova    & ID  &   rate           &  $N^{ISM}_H$                 & distance   &   distance\\ 
              & year    &     &   (\cps)          &  ($10^{21}$cm$^{-2}$)  &    (pc)   &   Ref. \\ 
      \noalign{\smallskip}\hline \hline\noalign{\smallskip}
       AT Cnc & 1700    &sm04 & 0.254$\pm$0.031  &  0.3     &  455$\pm$7             & 2 \\ \
       BD Pav & 1934    &sm04 & 0.423$\pm$0.034  &  0.5     &  329$\pm$2             & 2\\
       BT Mon & 1939    &sm04 & 0.024$\pm$0.009  &  3.0     & 1470$\pm$70            & 2 \\ 
       CG CMa & 1934    &em04 & 0.029$\pm$0.016  &  0.9     & 3500$\pm$500            & 2 \\ 
       CP Pup & 1942    &sm04 & 0.891$\pm$0.044  &  2.0     &  767$\pm$9               & 2 \\  
       DN Gem & 1912    &sm04 & 0.040$\pm$0.012  &  1.0     & 1330$\pm$90             & 2 \\  
       FM Cir & 2018    &em05 & 0.042$\pm$0.017  &  2.9     & 3200$\pm$700            & 2 \\  
       HV Vir & 1929    &sm04 & 0.038$\pm$0.011  &  0.2     &  320$\pm$25                & 1 \\  
       HZ Pup & 1963    &sm04 & 0.227$\pm$0.024  &  2.3     & 2800$\pm$500              & 2 \\  
       KT Eri & 2009    &sm04 & 0.066$\pm$0.012  &  1.0     & 4050$\pm$450              & 2 \\  
       RR Cha & 1953    &sm03 & 0.010$\pm$0.004  &  3.7     & 2200$\pm$800              & 2 \\  
       RR Pic & 1925    &sm04 & 0.280$\pm$0.010  &  0.1     &  492$\pm$5                 & 2 \\  
       T Sco  & 1860    &sm04 & 0.130$\pm$0.021  &  1.8     & 10340$\pm$120             & 5 \\  
   V1213 Cen  & 2009    &sm03 & 0.015$\pm$0.007  &  5.0     & 7200$^{+2400}_{-1800}$    & 7 \\  
   V1280 Sco  & 2007    &sm04 & 0.181$\pm$0.022  &  2.7     & 3300$\pm$200              & 2 \\ 
   V1313 Sco  & 2011    &sm04 & 0.018$\pm$0.007  &  9.0     & 5000$\pm$1500             & 1 \\  
   V1369 Cen  & 2013    &sm04 & 0.096$\pm$0.014  &  2.9     & 2400$^{+900}_{-500}$      & 8\\  
   V1706 Sco  & 2019    &sm04 & 0.063$\pm$0.014  &  11      & 6000$^{+3000}_{-2000}$    & 1 \\  
   V1708 Sco  & 2020    &sm04 & 5.159$\pm$0.110  &  14      & 6000$\pm$2000             & 1 \\  
   V1710 Sco  & 2021    &sm04 & 6.267$\pm$0.118  &  12      & 5000$\pm$2000             & 1 \\  
   V359 Cen   & 1939    &sm04 & 0.045$\pm$0.010  &  0.6     &  345$\pm$20                & 1 \\  
   V382 Vel   & 1999    &sm04 & 0.060$\pm$0.011  &  2.1     & 1520$\pm$90               & 2 \\  
   V407 Lup   & 2016    &sm04 & 0.125$\pm$0.016  &  2.5     &11000$\pm$3000            & 1 \\  
   V549 Vel   & 2017    &sm04 & 0.030$\pm$0.010  &  4.0     & 1400$\pm$400              & 1 \\  
   V598 Pup   & 2007    &sm04 & 0.031$\pm$0.008  &  1.2     & 1730$\pm$100              & 2  \\  
   V840 Oph   & 1917    &sm04 & 0.035$\pm$0.011  &  2.1     & 8000$\pm$800              & 6\\  
   V842 Cen   & 1986    &sm04 & 0.057$\pm$0.011  &  4.6     & 1325$\pm$90               & 2 \\  
   V888 Cen   & 1995    &sm04 & 0.017$\pm$0.006  &  4.8     & 2900$\pm$400              & 2 \\  
   V959 Mon   & 2012    &sm04 & 0.025$\pm$0.011  &   7      & 4000$\pm$1500             & 2 \\  
   YZ Ret     & 2020    &sm04 & 26.94$\pm$0.41   &   1      & 2525$\pm$250              & 2 \\ \hline
   \end{longtable}  
   \tablebib{
(1)~\citet{2021AJ....161..147B};
(2) \citet{2023AJ....165..163C}; (3) \citet{2002ApJ...574..950S}; (4) \citet{1988ASSL..145...47W};
(5) \citet{2021MNRAS.505.5957B}; (6) \citet{1996MNRAS.282..623H}; (7) \citet{2016Natur.537..649M};
(8) \citet{2021A&A...649A..28M}.
}
\end{center}

\begin{landscape}
\begin{longtable}{lccccccccc}
\caption{Properties of individual and merged detections \label{tab:detailed_summary} }\\
\hline \hline
        Name    & ID   & Pos. err.\footnotemark[5](\arcsec)& Sep(\arcsec)  & rate (\cps)     & counts      &   HR$_{P12}$   & HR$_{P23}$      & HR$_{P34}$      & Obs. date  \\ 
\hline
\endfirsthead
\caption{continued.}\\
\hline \hline
        Name    & ID   & Pos.err. (\arcsec) & Sep(\arcsec)  & rate (\cps)     & counts      &   HR$_{P12}$   & HR$_{P23}$      & HR$_{P34}$      & Obs. date  \\ 
\hline
\endhead
\hline
\endfoot
        AT Cnc & em01& 4.38 & 10   & 0.129$\pm$0.045 &   9.68$\pm$3.4 & 1.00$\pm$0.40  &  0.26$\pm$0.33  & -0.64$\pm$0.37  &   2020-04-25/26  \\
               & em02& 3.03 & 2.9  & 0.360$\pm$0.074 &   26.0$\pm$5.4 & 0.72$\pm$0.19  & -0.04$\pm$0.21  & -1.00$\pm$0.16  &   2020-10-28     \\
               & em03& 3.11 & 1.9  & 0.224$\pm$0.052 &   20.5$\pm$4.7 & 1.00$\pm$0.11  & -0.28$\pm$0.22  & -0.47$\pm$0.34  &   2021-04-28/29    \\
               & em04& 2.89 & 3.2  & 0.308$\pm$0.072 &   19.8$\pm$4.7 & 0.70$\pm$0.19  & -0.38$\pm$0.23  & -0.88$\pm$0.37  &   2021-10-30 \\
               & sm03& 1.97 & 1.0  & 0.242$\pm$0.033 &   57.1$\pm$7.9 & 0.88$\pm$0.09  & -0.09$\pm$0.14  & -0.72$\pm$0.16  &    - \\
               & sm04& 1.67 & 1.4  & 0.254$\pm$0.031 &   76.1$\pm$9.2 & 0.82$\pm$0.09  & -0.15$\pm$0.12  & -0.73$\pm$0.15  &    - \\
        BD Pav & em01& 2.13 & 3.1  & 0.415$\pm$0.072 &   40.2$\pm$6.9 & 0.89$\pm$0.11  &  0.01$\pm$0.18  & -0.35$\pm$0.22  &   2020-04-06/07    \\
               & em02& 1.61 & 2.3  & 0.611$\pm$0.088 &   53.5$\pm$7.7 & 0.81$\pm$0.19  &  0.56$\pm$0.13  & -0.39$\pm$0.15  &   2020-10-10/11    \\
               & em03& 2.35 & 2.2  & 0.204$\pm$0.045 &   23.4$\pm$5.1 & 0.62$\pm$0.21  & -0.39$\pm$0.23  & -0.22$\pm$0.39  &   2021-04-04/05    \\
               & em04& 1.89 & 2.1  & 0.499$\pm$0.065 &   66.8$\pm$8.7 & 0.57$\pm$0.34  &  0.80$\pm$0.09  & -0.34$\pm$0.12  &   2021-10-06/07    \\
               & sm03& 1.22 & 0.6  & 0.394$\pm$0.039 &   116$\pm$11  & 0.79$\pm$0.09  &  0.17$\pm$0.10  & -0.36$\pm$0.12   &     - \\
               & sm04& 1.08 & 1.0  & 0.423$\pm$0.034 &   181$\pm$14  & 0.75$\pm$0.10  &  0.39$\pm$0.08  & -0.34$\pm$0.09   &     - \\
        BT Mon & em01& 5.92 & 7.5  & 0.040$\pm$0.023 &    3.4$\pm$2.0 & 1.00$\pm$0.70  & -0.29$\pm$0.60  & -1.00$\pm$1.37  &   2020-04-10/11    \\
               & em04& 5.96 & 7.1  & 0.047$\pm$0.023 &    5.4$\pm$2.7 & 1.00$\pm$0.39  & -0.25$\pm$0.45  & -1.00$\pm$2.27  &   2021-10-11/12    \\
               & sm04& 4.14 & 5.6  & 0.024$\pm$0.009 &    9.1$\pm$3.5 & 1.00$\pm$0.22  & -0.34$\pm$0.34  & -1.00$\pm$1.37  &     - \\
        CG CMa & em04& 5.38 & 6.5  & 0.029$\pm$0.016 &    3.8$\pm$2.1 & 0.79$\pm$0.50  & -1.00$\pm$0.50  &                 &   2021-10-20/21    \\
        CP Pup & em01& 1.11 & 2.2  & 1.058$\pm$0.092 &   145$\pm$12  & 0.83$\pm$0.08  &  0.23$\pm$0.09  & -0.51$\pm$0.10   &   2020-05-06/07    \\
               & em02& 1.49 & 2.7  & 0.700$\pm$0.078 &    90$\pm$10  & 0.84$\pm$0.12  &  0.47$\pm$0.10  & -0.58$\pm$0.11   &   2020-11-07/08    \\
               & em03& 1.41 & 3.7  & 0.970$\pm$0.100 &   105$\pm$11  & 0.96$\pm$0.07  &  0.55$\pm$0.08  & -0.61$\pm$0.09   &   2021-05-09/10    \\
               & em04& 1.23 & 3.4  & 0.834$\pm$0.084 &   107$\pm$11  & 0.77$\pm$0.12  &  0.45$\pm$0.09  & -0.64$\pm$0.10   &   2021-11-09/10    \\
               & sm03& 0.86 & 2.5  & 0.913$\pm$0.052 &   337$\pm$20  & 0.86$\pm$0.06  &  0.39$\pm$0.06  & -0.55$\pm$0.06   &     - \\
               & sm04& 0.79 & 2.4  & 0.891$\pm$0.044 &   442$\pm$22  & 0.84$\pm$0.05  &  0.40$\pm$0.05  & -0.57$\pm$0.05   &     - \\
        DNGem  & em01& 6.12 & 4.3  & 0.042$\pm$0.025 &    3.2$\pm$1.9 & 1.00$\pm$1.42  &  0.38$\pm$0.54  & -1.00$\pm$0.83  &   2020-04-10/11   \\
               & em03& 4.34 & 1.7  & 0.054$\pm$0.025 &    5.1$\pm$2.3 & 1.00$\pm$0.70  &  0.02$\pm$0.51  & -0.27$\pm$0.63  &   2021-04-09/10   \\
               & sm03& 3.25 & 1.5  & 0.044$\pm$0.015 &   10.7$\pm$3.5 & 1.00$\pm$0.30  &  0.01$\pm$0.34  & -0.69$\pm$0.42  &     - \\
               & sm04& 2.80 & 1.9  & 0.040$\pm$0.012 &   13.9$\pm$4.1 & 1.00$\pm$0.25  &  0.14$\pm$0.29  & -0.84$\pm$0.33  &     - \\
        FM Cir & em05& 4.48 & 8.9  & 0.042$\pm$0.017 &    8.3$\pm$3.5 & 1.00$\pm$0.31  & -0.19$\pm$0.40  & -0.21$\pm$0.55  &    2022-02-15/17   \\
        HV Vir & em04& 4.94 & 9.7  & 0.062$\pm$0.026 &     7.0$\pm$2.9 & 0.04$\pm$0.40  & -1.00$\pm$0.74  &                &    2021-07-02/03    \\
               & em05& 5.12 & 1.5  & 0.077$\pm$0.030 &     8.7$\pm$3.4 & 1.00$\pm$0.36  & -0.61$\pm$0.39  &  0.00$\pm$0.79 &    2022-01-03/04    \\
               & sm03& 4.28 & 9.2  & 0.032$\pm$0.011 &    11.1$\pm$4.0 & 1.00$\pm$0.22  & -0.26$\pm$0.36  & -0.11$\pm$0.57 &     - \\
               & sm04& 3.22 & 6.6  & 0.038$\pm$0.011 &    17.0$\pm$4.8 & 0.67$\pm$0.30  & -0.42$\pm$0.31  & -0.16$\pm$0.62 &     - \\
        HZ Pup & em01& 2.69 & 2.4  & 0.288$\pm$0.055 &    30.9$\pm$5.9 & 0.48$\pm$0.20  & -0.51$\pm$0.19  &  0.29$\pm$0.27 &    2020-05-01/02    \\
               & em02& 5.08 & 5.1  & 0.042$\pm$0.022 &     4.8$\pm$2.5 & 1.00$\pm$0.46  & -0.50$\pm$0.47  &  0.60$\pm$0.39 &    2020-11-02/03    \\
               & em03& 2.09 & 3.3  & 0.402$\pm$0.061 &    48.8$\pm$7.4 & 0.39$\pm$0.19  &  0.06$\pm$0.17  & -0.16$\pm$0.20 &    2021-05-05/06   \\
               & em04& 3.25 & 2.2  & 0.149$\pm$0.041 &    16.8$\pm$4.6 & 0.54$\pm$0.46  &  0.29$\pm$0.31  & -0.32$\pm$0.33 &    2021-11-04/05    \\
               & sm03& 1.56 & 2.0  & 0.249$\pm$0.029 &    84.4$\pm$9.7 & 0.49$\pm$0.13  & -0.19$\pm$0.13  &  0.05$\pm$0.16 &     - \\
               & sm04& 1.42 & 0.9  & 0.227$\pm$0.024 &    102$\pm$11  & 0.49$\pm$0.12   & -0.11$\pm$0.12  & -0.01$\pm$0.14 &     - \\   
        KT Eri & em01& 3.28 & 6.1  & 0.083$\pm$0.026 &    12.5$\pm$3.9 & 0.27$\pm$0.60  &  0.64$\pm$0.25  & -0.53$\pm$0.32 &   2020-03-09/10     \\
               & em02& 4.97 & 7.9  & 0.058$\pm$0.022 &     7.9$\pm$3.0 & 0.77$\pm$0.42  & -0.14$\pm$0.39  & -0.74$\pm$0.71 &   2020-09-08/09     \\
               & em03& 4.11 & 1.8  & 0.060$\pm$0.022 &     9.3$\pm$3.5 & 0.32$\pm$0.57  &  0.18$\pm$0.46  &  0.17$\pm$0.42 &   2021-02-25/27     \\
               & em04& 4.32 & 3.5  & 0.059$\pm$0.022 &     9.3$\pm$3.5 & 1.00$\pm$0.45  &  0.04$\pm$0.38  & -0.21$\pm$0.55 &   2021-09-01/02    \\
               & sm03& 2.49 & 3.7  & 0.068$\pm$0.014 &    29.7$\pm$6.0  & 0.50$\pm$0.30  &  0.32$\pm$0.21  & -0.34$\pm$0.26&     - \\
               & sm04& 2.15 & 3.1  & 0.066$\pm$0.012 &    39.0$\pm$6.9  & 0.63$\pm$0.25  &  0.25$\pm$0.19  & -0.28$\pm$0.24&     -\\
        RR Cha & sm03& 5.20 & 1.2  & 0.010$\pm$0.004 &     7.1$\pm$3.3 & 0.83$\pm$0.48  &  0.81$\pm$0.52  & -0.39$\pm$0.78 &     - \\
        RR Pic & em01& 1.09 & 1.9  & 0.253$\pm$0.019 &    211$\pm$16  & 0.57$\pm$0.07  & -0.35$\pm$0.08  & -0.69$\pm$0.12  &   2020-05-03/12     \\
               & em02& 0.97 & 3.8  & 0.296$\pm$0.020 &    238$\pm$16  & 0.53$\pm$0.07  & -0.29$\pm$0.07  & -0.63$\pm$0.11  &   2020-11-03/11    \\
               & em03& 1.13 & 2.0  & 0.264$\pm$0.021 &    182$\pm$14  & 0.63$\pm$0.08  & -0.22$\pm$0.08  & -0.62$\pm$0.12  &   2021-05-07/14     \\
               & em04& 1.10 & 3.1  & 0.290$\pm$0.021 &    223$\pm$16  & 0.56$\pm$0.07  & -0.28$\pm$0.08  & -0.76$\pm$0.10  &   2021-11-05/13     \\
               & sm03& 0.75 & 1.6  & 0.274$\pm$0.012 &    631$\pm$27  & 0.58$\pm$0.04  & -0.29$\pm$0.04  & -0.67$\pm$0.07  &     - \\
               & sm04& 0.70 & 2.1  & 0.280$\pm$0.010 &    856$\pm$31  & 0.57$\pm$0.04  & -0.30$\pm$0.04  & -0.69$\pm$0.06  &     - \\
        T Sco  & em02& 4.02 & 4.0  & 0.094$\pm$0.031 &    11.4$\pm$3.8 & 0.21$\pm$0.51  &  0.25$\pm$0.39  & -0.57$\pm$0.46 &    2020-09-05/06   \\
               & sm03& 3.24 & 5.2  & 0.084$\pm$0.017 &    33.5$\pm$6.8 & 0.86$\pm$0.22  &  0.28$\pm$0.21  & -0.78$\pm$0.19 &     - \\
               & sm04& 4.22 & 4.5  & 0.130$\pm$0.021 &    69$\pm$11  & 0.78$\pm$0.18  &  0.06$\pm$0.17  & -0.47$\pm$0.21   &     - \\
        TV Crv & em01& 4.30 & 1.0  & 0.066$\pm$0.025 &    8.6$\pm$3.3 & 0.50$\pm$0.43  & -0.10$\pm$0.41  & -1.00$\pm$0.58  &    2019-12-22/23   \\
               & em03& 5.22 & 2.6  & 0.032$\pm$0.019 &    3.5$\pm$2.0 & 1.00$\pm$0.40  & -0.43$\pm$0.50  & -1.00$\pm$1.54  &    2020-12-23/24  \\
               & em05& 3.62 & 2.0  & 0.134$\pm$0.038 &    15.1$\pm$4.3 & 0.94$\pm$0.24  & -0.05$\pm$0.30  & -1.00$\pm$0.65 &    2021-12-28/29   \\
               & sm03& 2.89 & 3.2  & 0.051$\pm$0.015 &    17.8$\pm$5.4 & 0.62$\pm$0.27  & -0.19$\pm$0.28  & -1.00$\pm$0.41 &     - \\
               & sm04& 2.79 & 2.6  & 0.057$\pm$0.013 &    25.3$\pm$5.8 & 0.54$\pm$0.30  &  0.06$\pm$0.26  & -0.69$\pm$0.29 &     - \\
    V1213 Cen  & sm03& 4.91 & 6.8  & 0.015$\pm$0.007 &     8.5$\pm$3.8 & 1.00$\pm$0.36  &  0.22$\pm$0.46  & -0.94$\pm$0.42 &     - \\
    V1280 Sco  & em01& 2.36 & 2.6  & 0.252$\pm$0.050 &     30.1$\pm$6.0 & 0.57$\pm$0.17  & -0.67$\pm$0.17  & -1.00$\pm$0.51&    2020-03-18/19    \\
               & em02& 2.79 & 1.7  & 0.153$\pm$0.038 &     20.1$\pm$5.0 & 0.49$\pm$0.20  & -1.00$\pm$0.07  &               &    2020-09-18/19   \\
               & em03& 2.68 & 1.1  & 0.222$\pm$0.053 &     21.5$\pm$5.1 & 0.62$\pm$0.19  & -0.91$\pm$0.13  & -1.00$\pm$2.38&    2021-03-07/15   \\
               & em04& 3.54 & 2.9  & 0.126$\pm$0.035 &     16.2$\pm$4.5 & 0.10$\pm$0.30  & -0.92$\pm$0.24  & -1.00$\pm$6.81&    2021-09-13/14    \\
               & sm03& 1.57 & 0.6  & 0.205$\pm$0.027 &     70.4$\pm$9.3 & 0.55$\pm$0.11  & -0.86$\pm$0.08  & -1.00$\pm$0.46&     - \\
               & sm04& 1.48 & 0.9  & 0.181$\pm$0.022 &     85$\pm$10  & 0.43$\pm$0.11  & -0.85$\pm$0.08  & -1.00$\pm$0.39  &     - \\
    V1313 Sco  & sm04& 4.24 & 7.5  & 0.018$\pm$0.007 &     8.5$\pm$3.5 &               &  1.00$\pm$1.34  &  0.41$\pm$0.32  &     - \\ 
    V1369 Cen  & em01& 2.81 & 2.1  & 0.113$\pm$0.028 &     20.3$\pm$5.0 & 0.44$\pm$0.27  & -0.02$\pm$0.26  & -0.95$\pm$0.28&    2020-02-18/21   \\
               & em02& 3.43 & 5.5  & 0.098$\pm$0.025 &     19.4$\pm$5.0 & 0.24$\pm$0.32  & -0.03$\pm$0.30  & -0.47$\pm$0.35&    2020-08-16/18    \\
               & em03& 3.87 & 5.9  & 0.086$\pm$0.029 &     11.2$\pm$3.7 & 0.82$\pm$0.26  & -0.60$\pm$0.30  & -0.25$\pm$0.75&    2021-02-02/04    \\
               & em04& 2.98 & 3.2  & 0.088$\pm$0.026 &     13.5$\pm$4.0 & 0.42$\pm$0.39  &  0.19$\pm$0.31  & -1.00$\pm$0.27&    2021-08-11/12    \\
               & em05& 3.38 & 4.0  & 0.096$\pm$0.027 &     15.8$\pm$4.5 & 0.65$\pm$0.25  & -0.45$\pm$0.28  & -0.50$\pm$0.61&    2022-02-09/10     \\
               & sm03& 2.16 & 2.7  & 0.099$\pm$0.016 &     49.8$\pm$7.9 & 0.49$\pm$0.17  & -0.19$\pm$0.17  & -0.57$\pm$0.25&     - \\
               & sm04& 1.79 & 2.6  & 0.096$\pm$0.014 &     62.3$\pm$8.8 & 0.48$\pm$0.15  & -0.11$\pm$0.15  & -0.72$\pm$0.20&     - \\
    V1706 Sco  & em02& 3.37 & 1.6  & 0.092$\pm$0.030 &     11.9$\pm$3.9 & 1.00$\pm$1.56  &  0.82$\pm$0.27  & -0.72$\pm$0.31&    2020-09-20/22    \\
               & em03& 4.88 & 2.2  & 0.097$\pm$0.038 &     9.57$\pm$3.7 & 1.00$\pm$0.69  &  0.16$\pm$0.40  & -1.00$\pm$0.42&    2021-03-17/18    \\
               & sm03& 2.63 & 2.2  & 0.068$\pm$0.017 &     23.0$\pm$5.7 & 1.00$\pm$0.25  &  0.45$\pm$0.24  & -0.72$\pm$0.29&     - \\
               & sm04& 2.64 & 4.2  & 0.063$\pm$0.014 &     29.0$\pm$6.5 & 0.74$\pm$0.26  &  0.35$\pm$0.24  & -0.61$\pm$0.30&     - \\
    V1708 Sco  & em02& 0.58 & 0.9  & 17.905$\pm$0.382 &    2340$\pm$50  & 0.98$\pm$0.01  & -0.36$\pm$0.05  & -0.82$\pm$0.02 &    2020-09-24/25   \\
               & sm03& 0.57 & 0.2 & 7.133$\pm$0.152 &     2330$\pm$50  & 0.98$\pm$0.01  & -0.39$\pm$0.03  & -0.80$\pm$0.03&     - \\
               & sm04& 0.57 & 0.6 & 5.159$\pm$0.110 &     2330$\pm$50  & 0.98$\pm$0.01  & -0.40$\pm$0.02  & -0.78$\pm$0.03&     - \\
    V1710 Sco  & em04& 0.56 & 2.0  & 22.914$\pm$0.429 &    2990$\pm$60  & 0.37$\pm$0.02  & -0.94$\pm$0.01  & -0.83$\pm$0.10 &    2021-09-17/18   \\
               & sm04& 0.55 & 1.8  & 6.267$\pm$0.118 &     2980$\pm$60  & 0.37$\pm$0.02  & -0.94$\pm$0.01  & -0.79$\pm$0.11&      - \\
    V359 Cen   & em01& 5.04 & 6.4  & 0.029$\pm$0.015 &      4.9$\pm$2.6 &                &  1.00$\pm$0.24  & -1.00$\pm$0.26&    2019-12-30/31    \\
               & em02& 3.90 & 2.6  & 0.084$\pm$0.027 &     11.5$\pm$3.7 & 0.58$\pm$0.72  &  0.89$\pm$0.21  & -1.00$\pm$0.21&    2020-06-29/30    \\
               & em03& 4.92 & 4.1  & 0.051$\pm$0.022 &      7.1$\pm$3.0 & 0.13$\pm$0.86  &  0.59$\pm$0.49  & -0.60$\pm$0.60&    2020-12-31      \\
               & em05& 5.51 & 2.3  & 0.032$\pm$0.017 &      4.8$\pm$2.5 & 1.00$\pm$0.24  & -1.00$\pm$0.25  &               &    2022-01-04      \\
               & sm03& 3.40 & 0.9 & 0.051$\pm$0.012 &     22.6$\pm$5.4 & 0.43$\pm$0.65  &  0.88$\pm$0.17  & -1.00$\pm$0.19&     - \\
               & sm04& 3.38 & 3.5  & 0.045$\pm$0.010 &     26.5$\pm$5.9 & 0.25$\pm$0.45  &  0.63$\pm$0.19  & -1.00$\pm$0.15&     - \\
    V382 Vel   & em01& 4.50 & 5.2  & 0.042$\pm$0.017 &     7.5$\pm$3.0 & 1.00$\pm$0.55  &  0.28$\pm$0.38  & -0.65$\pm$0.42 &    \\
               & em03& 3.64 & 4.8  & 0.055$\pm$0.019 &     10.1$\pm$3.4 &                &  1.00$\pm$0.13  & -0.39$\pm$0.29&    2020-12-24/26    \\
               & em04& 3.97 & 1.4  & 0.128$\pm$0.030 &     20.9$\pm$5.0 & 0.77$\pm$0.29  &  0.38$\pm$0.23  & -0.25$\pm$0.25&    2021-06-26/28    \\
               & em05& 3.48 & 0.7  & 0.063$\pm$0.021 &     11.5$\pm$3.7 & 1.00$\pm$0.84  &  0.50$\pm$0.29  & -0.51$\pm$0.38&    2021-12-28/30    \\
               & sm03& 2.81 & 0.8  & 0.039$\pm$0.010 &     20.1$\pm$5.1 & 1.00$\pm$0.39  &  0.69$\pm$0.18  & -0.55$\pm$0.23&     - \\
               & sm04& 2.22 & 1.4  & 0.060$\pm$0.011 &     40.6$\pm$7.1 & 0.96$\pm$0.19  &  0.55$\pm$0.14  & -0.43$\pm$0.17&     - \\
    V407 Lup   & em01& 3.17 & 4.0  & 0.186$\pm$0.044 &     21.9$\pm$5.2 & 0.98$\pm$0.31  &  0.30$\pm$0.24  & -0.37$\pm$0.25&    2020-02-29/03-01    \\
               & em02& 2.69 & 3.4  & 0.159$\pm$0.035 &     24.6$\pm$5.4 & 1.00$\pm$0.24  &  0.46$\pm$0.21  & -0.52$\pm$0.23&    2020-08-30/31     \\
               & em03& 3.13 & 3.2  & 0.091$\pm$0.026 &     14.9$\pm$4.2 & 0.67$\pm$0.33  &  0.22$\pm$0.30  & -0.02$\pm$0.31&    2021-02-14/16     \\
               & em04& 3.72 & 1.7  & 0.080$\pm$0.028 &     10.3$\pm$3.6 & 1.00$\pm$0.31  &  0.03$\pm$0.35  & -0.65$\pm$0.45&    2021-08-23/25     \\
               & sm03& 1.85 & 2.4  & 0.140$\pm$0.020 &     61.3$\pm$8.6 & 0.86$\pm$0.14  &  0.36$\pm$0.14  & -0.34$\pm$0.15&     - \\
               & sm04& 1.73 & 1.8  & 0.125$\pm$0.016 &     70.9$\pm$9.3 & 0.89$\pm$0.12  &  0.32$\pm$0.13  & -0.37$\pm$0.14&     - \\
    V549 Vel   & sm03& 4.41 & 4.5  & 0.037$\pm$0.013 &     16.1$\pm$5.7 & 0.16$\pm$0.34  & -0.56$\pm$0.45  & -1.00$\pm$1.21&     - \\
               & sm04& 3.86 & 2.7  & 0.030$\pm$0.010 &     18.4$\pm$6.3 & 0.03$\pm$0.34  & -0.85$\pm$0.29  & -1.00$\pm$2.94&     - \\
    V598 Pup   & em04& 3.32 & 1.1  & 0.066$\pm$0.021 &     10.8$\pm$3.5 & 1.00$\pm$1.45  &  0.80$\pm$0.20  & -0.39$\pm$0.32&    2021-10-24/25     \\
               & sm03& 4.01 & 8.7  & 0.019$\pm$0.007 &      8.8$\pm$3.4 & 0.54$\pm$0.52  & -0.11$\pm$0.46  &  0.41$\pm$0.37&     - \\
               & sm04& 2.99 & 3.7  & 0.031$\pm$0.008 &     19.5$\pm$5.0 & 0.78$\pm$0.43  &  0.47$\pm$0.24  & -0.03$\pm$0.26&     - \\
    V840 Oph   & em01& 4.87 & 2.8  & 0.055$\pm$0.025 &      6.7$\pm$3.1 & 1.00$\pm$0.19  & -0.80$\pm$0.32  &  0.63$\pm$0.57&    2020-03-17/18    \\
               & em02& 4.68 & 5.9  & 0.062$\pm$0.027 &      8.2$\pm$3.6 & 1.00$\pm$0.38  &  0.01$\pm$0.44  & -0.17$\pm$0.50&    2020-09-16/18    \\
               & sm03& 3.56 & 3.4  & 0.046$\pm$0.014 &     15.8$\pm$5.0 & 1.00$\pm$0.11  & -0.17$\pm$0.31  &  0.02$\pm$0.36&     - \\
               & sm04& 3.61 & 1.7  & 0.035$\pm$0.011 &     16.4$\pm$5.3 & 1.00$\pm$0.12  &  0.01$\pm$0.30  & -0.04$\pm$0.32&     - \\
    V842 Cen   & em01& 6.07 & 10   & 0.076$\pm$0.027 &     10.5$\pm$3.7 & 1.00$\pm$0.89  &  0.68$\pm$0.28  & -0.34$\pm$0.32&    2020-02-25/26     \\
               & em02& 3.80 & 1.6  & 0.066$\pm$0.021 &     12.1$\pm$3.9 & 1.00$\pm$0.43  &  0.46$\pm$0.28  & -1.00$\pm$0.39&    2020-08-24/25     \\
               & em03& 4.32 & 5.5  & 0.053$\pm$0.021 &     7.9$\pm$3.1 & 0.18$\pm$0.73  &  0.53$\pm$0.40  & -0.20$\pm$0.46 &    2021-02-08/10     \\
               & em04& 5.33 & 3.6  & 0.037$\pm$0.018 &     5.7$\pm$2.7 & 1.00$\pm$0.86  &  0.43$\pm$0.41  & -0.52$\pm$0.49 &    2021-08-18/19     \\
               & sm03& 2.63 & 2.6  & 0.064$\pm$0.013 &     30.1$\pm$6.1 & 0.83$\pm$0.28  &  0.53$\pm$0.19  & -0.40$\pm$0.22&     - \\
               & sm04& 2.43 & 2.8  & 0.057$\pm$0.011 &     35.1$\pm$6.6 & 0.89$\pm$0.20  &  0.49$\pm$0.18  & -0.41$\pm$0.20&     - \\
    V888 Cen   & em01& 6.74 & 3.6  & 0.029$\pm$0.013 &      6.7$\pm$3.1 & 1.00$\pm$0.61  &  0.29$\pm$0.48  & -1.00$\pm$0.82&    2020-02-04/06     \\
               & sm03& 4.71 & 4.3  & 0.017$\pm$0.007 &      9.7$\pm$3.9 & 1.00$\pm$2.27  &  0.84$\pm$0.33  & -1.00$\pm$0.41&     - \\
               & sm04& 3.69 & 4.1  & 0.017$\pm$0.006 &     12.9$\pm$4.4 & 1.00$\pm$5.21  &  0.95$\pm$0.25  & -1.00$\pm$0.27&     - \\
    V959 Mon   & sm04& 4.73 & 6.5  & 0.025$\pm$0.011 &      8.7$\pm$3.8 & 0.53$\pm$0.42  & -0.74$\pm$0.45  &  0.67$\pm$0.57&     - \\
    YZ Ret     & em01& 4.87 & 6.8  & 0.043$\pm$0.013 &     15.8$\pm$4.8 & 0.92$\pm$0.28  & -0.01$\pm$0.30  & -0.73$\pm$0.49&    2020-01-05/09    \\
               & em03& 0.66 & 0.65 & 3.054$\pm$0.104 &     914$\pm$31  & 0.94$\pm$0.01  & -0.73$\pm$0.13  & -1.00$\pm$0.65 &    2020-07-04/08     \\
               & em04& 0.80 & 1.3 & 1.175$\pm$0.053 &     515$\pm$24  & 0.77$\pm$0.03  & -0.31$\pm$0.11  & -0.61$\pm$0.17 &    2021-01-04/08    \\
               & em05& 1.21 & 2.6  & 0.410$\pm$0.035 &     152$\pm$13  & 0.43$\pm$0.09  & -0.18$\pm$0.14  & -0.36$\pm$0.18 &    2021-07-09/10    \\
               & sm03& 1.44 & 4.8  & 35.39$\pm$1.10  &     41620$\pm$1300& 0.87$\pm$0.01  & -0.86$\pm$0.02  & -0.54$\pm$0.1&     - \\
               & sm04& 1.33 & 3.0  & 26.94$\pm$0.41  &     43060$\pm$660 & 0.86$\pm$0.01  & -0.82$\pm$0.02  & -0.56$\pm$0.1&     - \\ \hline
\end{longtable}
\footnotetext{Position error of source in the eROSITA catalogues, see sect. 6.2 in \citet{2024A&A...682A..34M}}
\end{landscape}

\section{Confidence contours for AT~Cnc and BD~Pav spectral fits}

\begin{figure}[!ht]
    \includegraphics[width=0.5\textwidth]{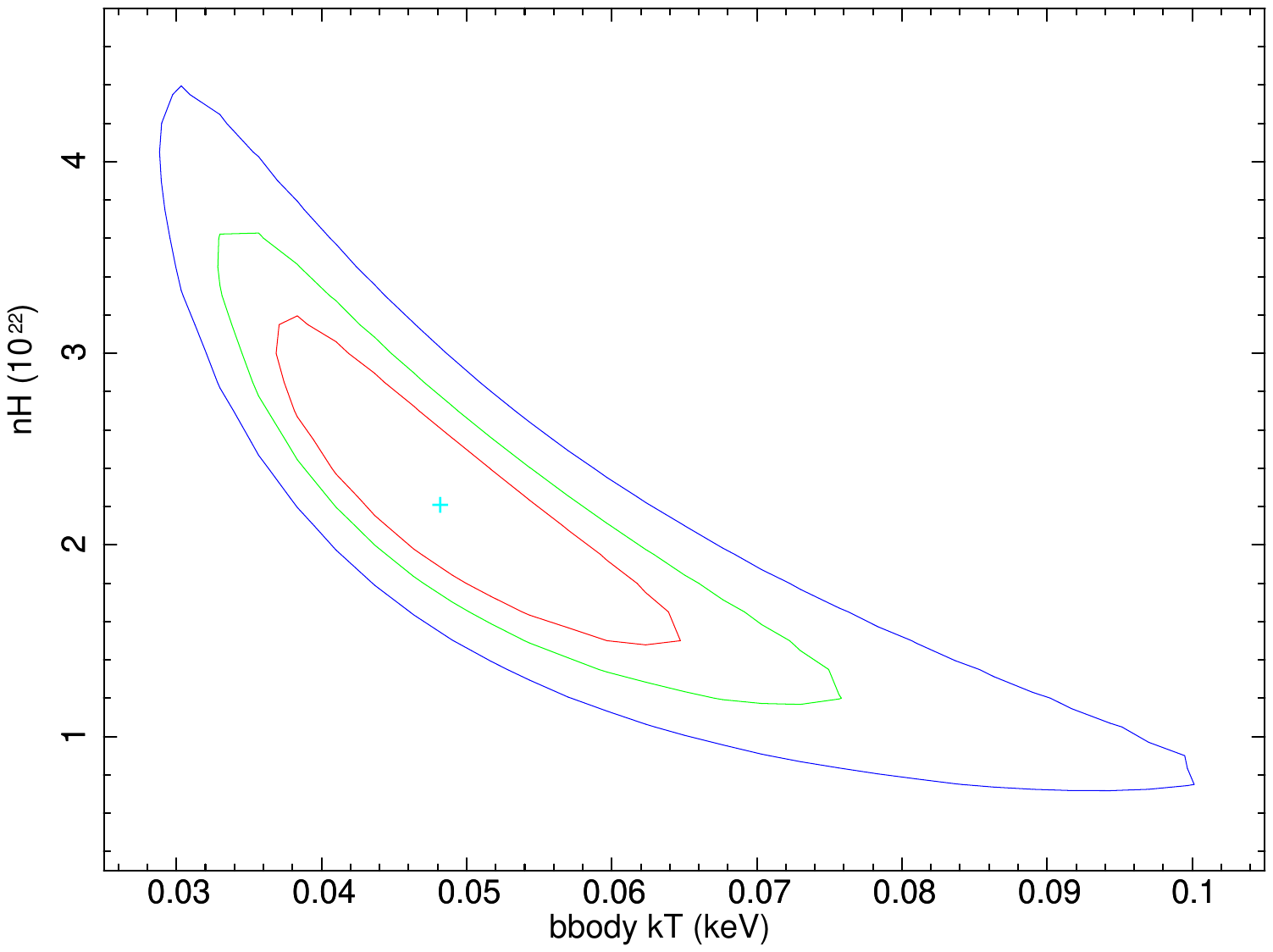}\hfill
    \includegraphics[width=0.53\textwidth]{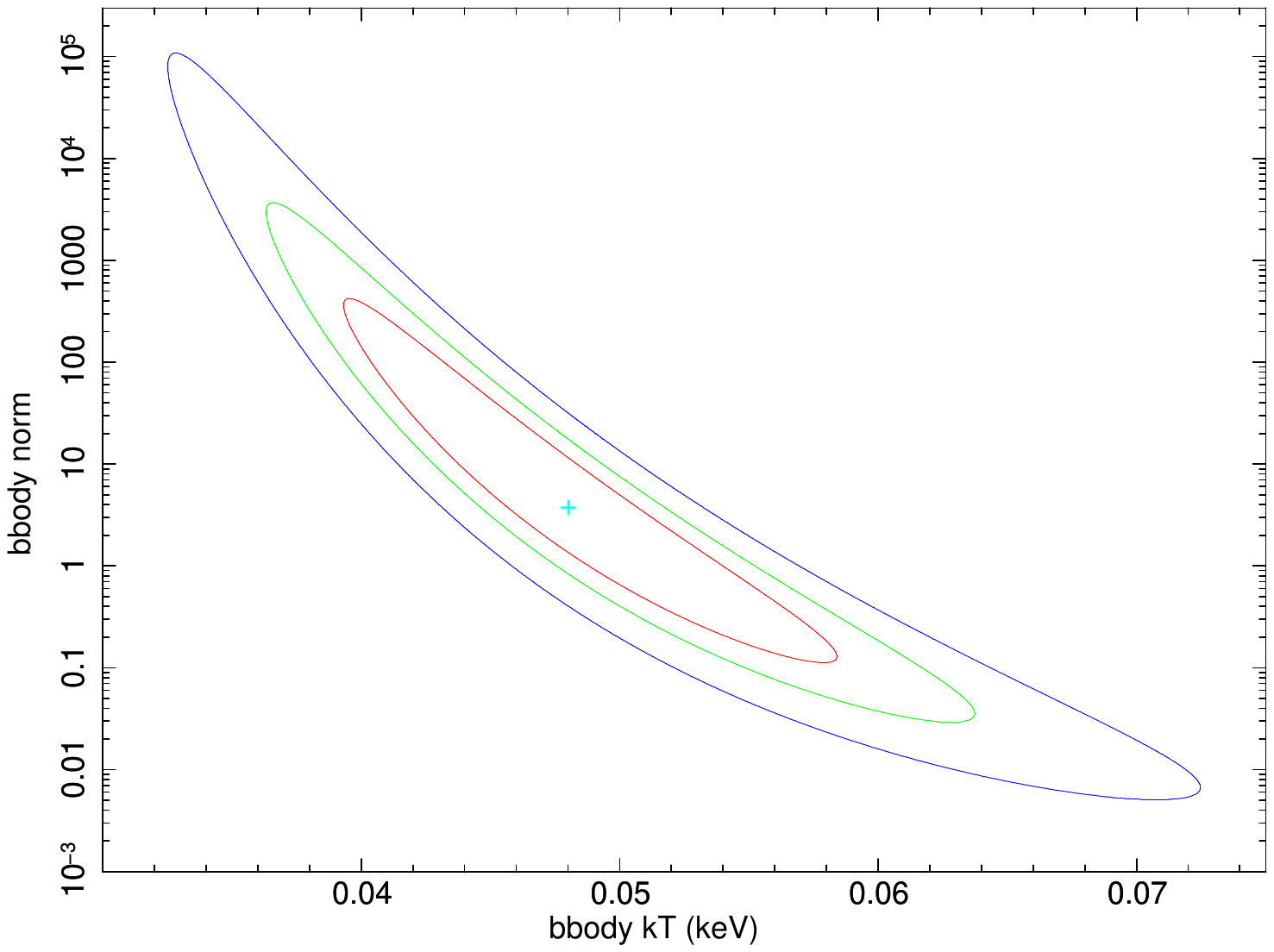}\hfill
    \includegraphics[width=0.5\textwidth]{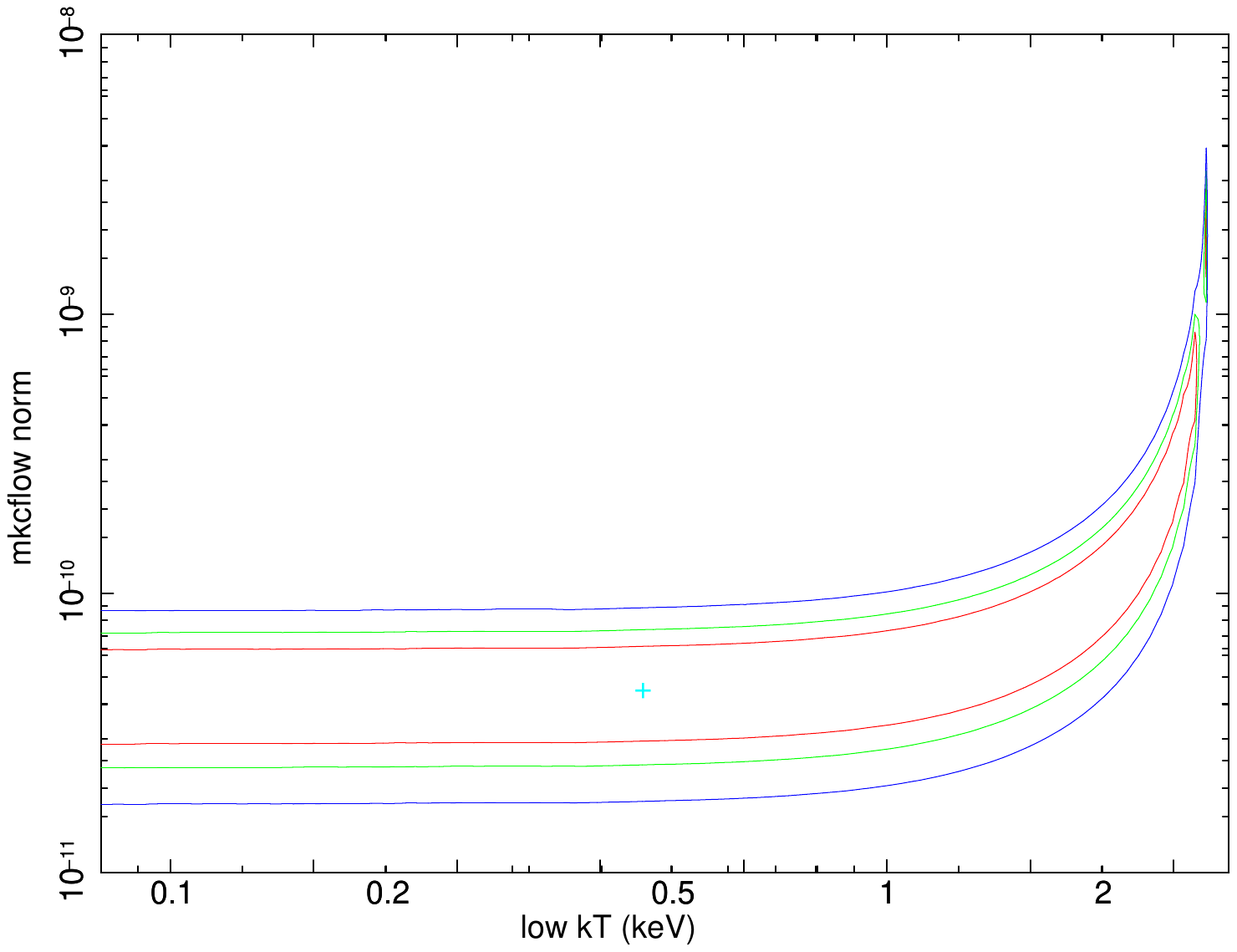}\hfill
    \includegraphics[width=0.5\textwidth]{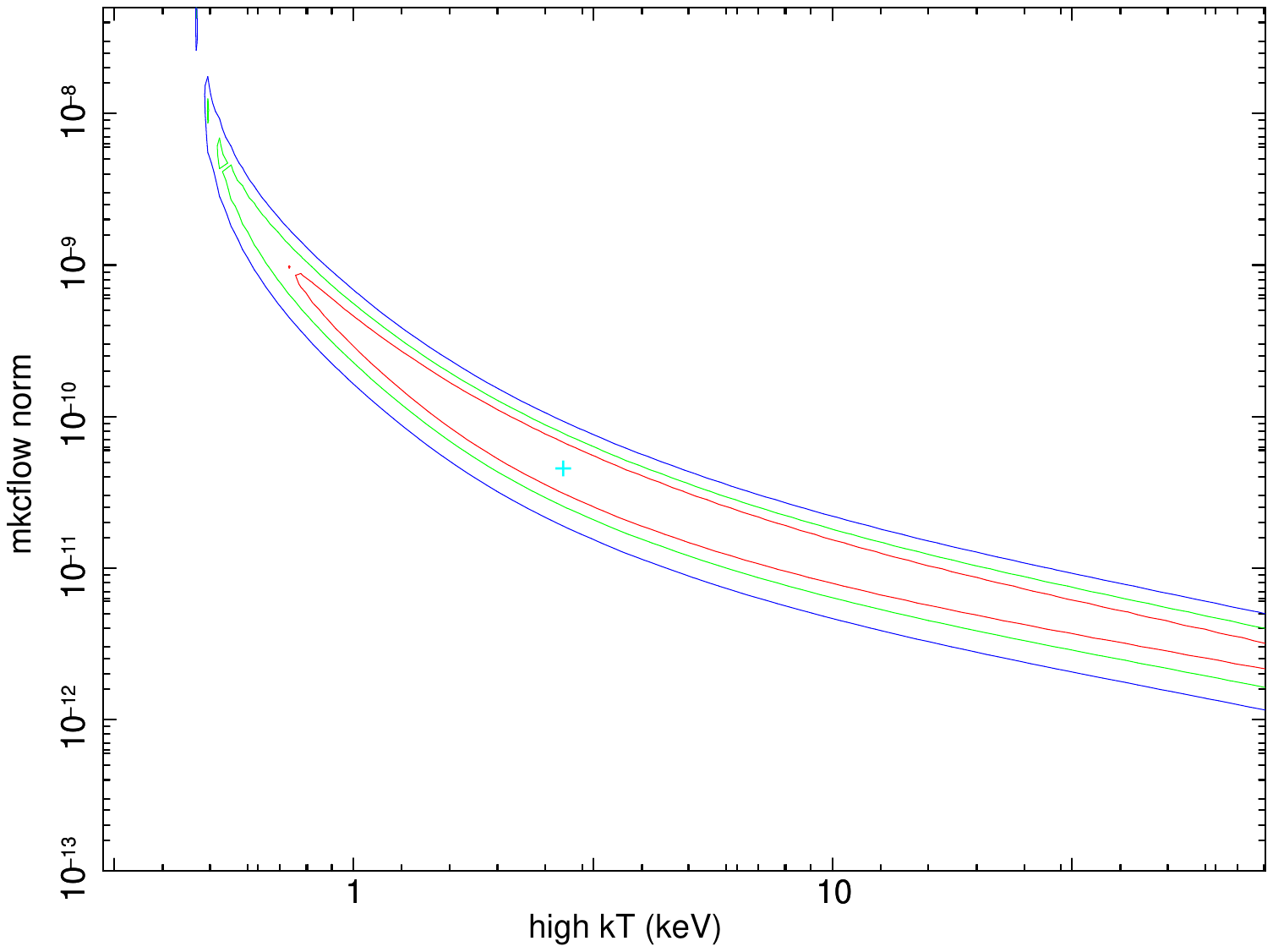}\hfill
    \caption{AT Cnc contour plots for the spectral model parameters at 68\%, 90\% and 99\% confidence level. From top left to bottom right: absorbing hydrogen column (in units of 10$^{22}$cm$^{-2}$) vs blackbody $kT$; blackbody normalization ( $L_{39}/D^2_{10}$, where $L_{39}$ is luminosity in units of $10^{39}$erg/s and $D_{10}$ is the distance in units of 10~kpc) vs blackbody $kT$; and cooling flow normalization (mass accretion rate in units of $M_{\odot}yr^{-1}$) vs lowest and highest plasma temperature of the cooling flow.
    \label{ATCnc_cont}}
\end{figure}

\begin{figure}[!ht]
    \includegraphics[width=0.5\textwidth]{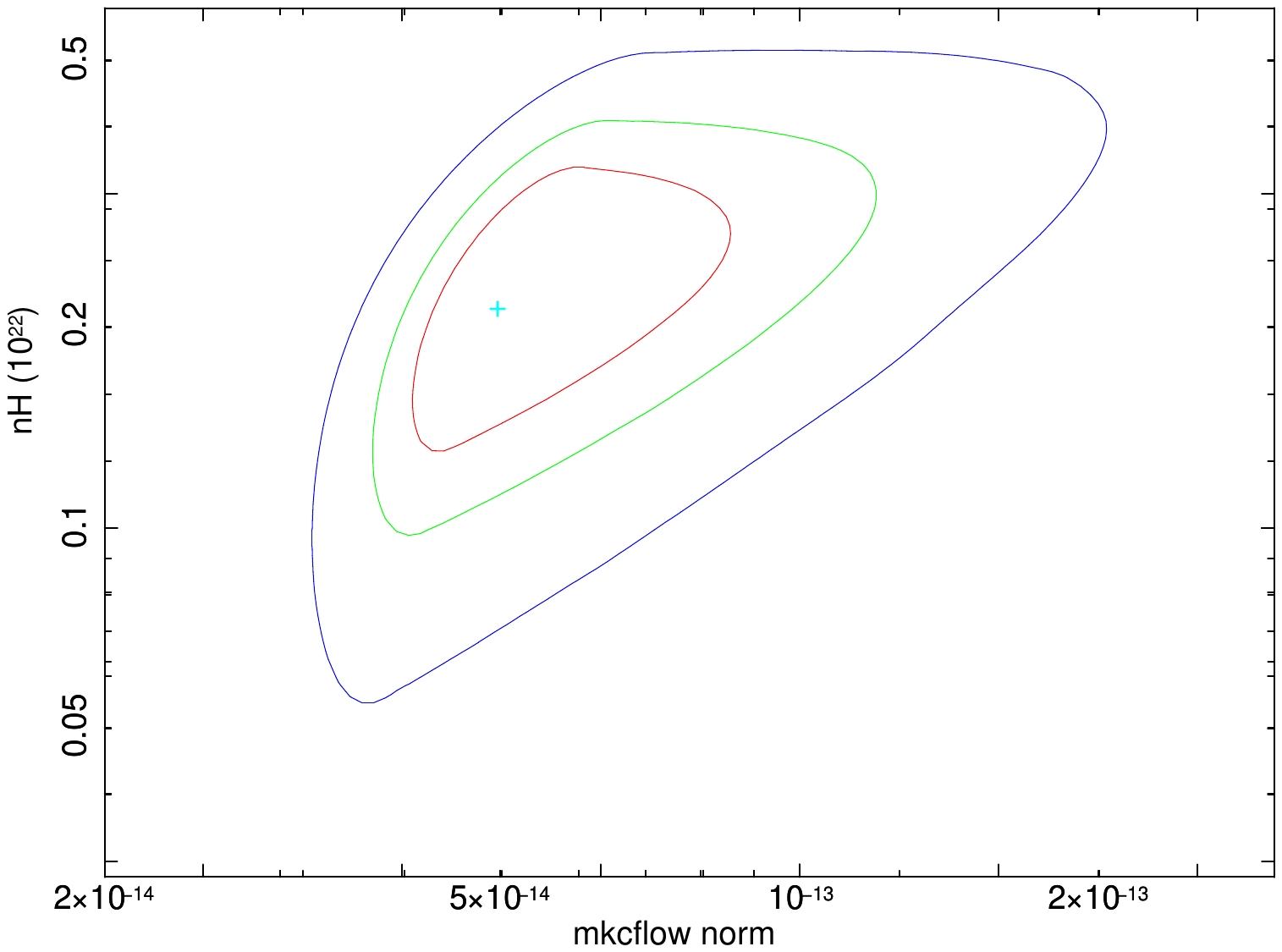}\hfill
    \includegraphics[width=0.5\textwidth]{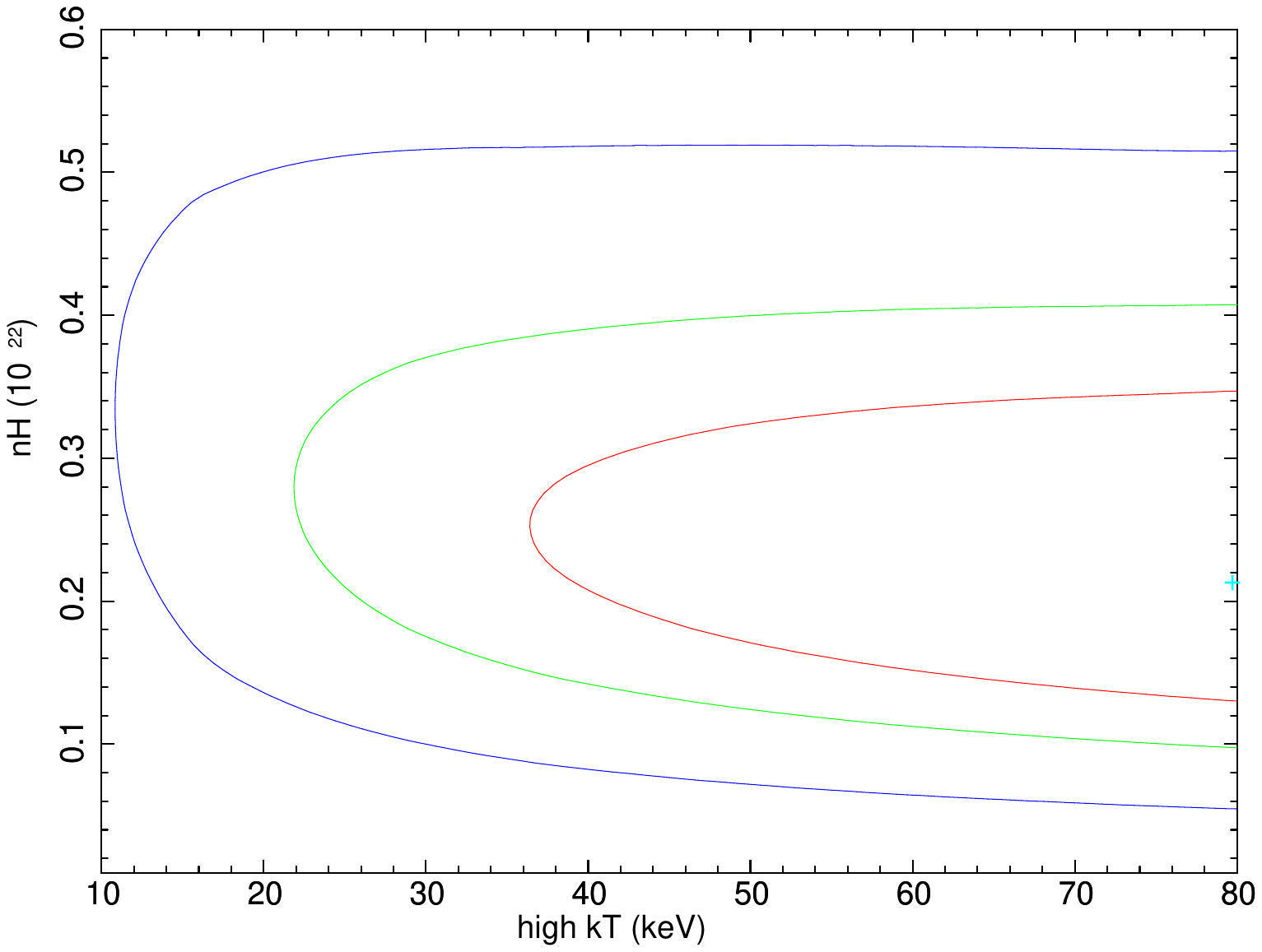}\hfill
    \caption{BD Pav countour plots for the spectral model parameters at 68\%, 90\% and 99\% confidence level. Absorbing hydrogen column (in units of 10$^{22}$cm$^{-2}$) vs cooling flow normalization (mass accretion rate in units of $M_{\odot}yr^{-1}$, left panel) and highest plasma temperature of the cooling flow (right panel).
\label{BDPav_spec_cont}}
\end{figure}

\end{appendix}

\end{document}